\documentclass[twocolumn,showpacs,prd,aps,floatfix,superscriptaddress,showkeys,citeautoscript,longbibliography,nobalancelastpage]{revtex4-2}
\usepackage[colorlinks=true,linkcolor=blue,citecolor=blue,urlcolor=blue]{hyperref}
\usepackage{graphicx,wrapfig,epstopdf,booktabs,soul,array,lmodern}
\usepackage[left=1.10cm, right=1.10cm, top=1.70cm, bottom=1.70cm]{geometry}
\usepackage{balance}
\usepackage{epigraph}
\usepackage{footmisc}
\usepackage{dcolumn}
\usepackage{dutchcal}
\usepackage{bm,color}
\usepackage{amsmath,amssymb,dsfont,amstext,amsfonts}
\usepackage{extarrows}
\usepackage{lettrine}
\usepackage{times,mathptmx}
\usepackage[english]{babel}
\usepackage{mathtools}
\usepackage{bbold}
\usepackage{wasysym}
\usepackage{mathtools}
\usepackage{bbold} 
\usepackage{braket}
\usepackage{dsfont}
\usepackage{bigints}
\usepackage{soul}
\usepackage{amssymb,amsmath,amsthm,amssymb,multirow,printlen,layouts}
\usepackage{color}
\definecolor{lightgray}{gray}{0.5}
\usepackage[export]{adjustbox}
\usepackage{footnote}
\usepackage{ucs}
\usepackage{braket}
\usepackage{dsfont}
\usepackage{bigints}
\usepackage[dvipsnames,svgnames,table]{xcolor}

\usepackage{soul}
\usepackage{url,hyperref}
\usepackage{caption}
\usepackage{charter}
\usepackage[T1]{fontenc}
\definecolor{CustomBlue}{rgb}{0.2, 0.18, 0.65} % Adjust the RGB values to match the desired blue
\usepackage{xcolor}
\definecolor{mydarkred}{rgb}{0.7, 0.0, 0.0}  % This is a dark blue
\hypersetup{
    backref=true,
    pagebackref=true,
    colorlinks=true,
    linkcolor=CustomBlue,
    citecolor=CustomBlue,
    urlcolor=CustomBlue
}
\DeclareUnicodeCharacter{2113}{\ensuremath{\ell}}

\usepackage{float}
\usepackage{fancyhdr}
\usepackage{fnpos}
\addto{\captionsenglish}{%
  
}
\usepackage[raggedright]{titlesec}
\usepackage{epstopdf}
\definecolor{cream}{RGB}{222,217,201}%\usepackage{setspace}

\usepackage{float}
\usepackage{fancyhdr}
\usepackage{orcidlink}
\usepackage{fnpos}
\addto{\captionsenglish}{%
  
}
\usepackage[raggedright]{titlesec}
\usepackage{epstopdf}
\definecolor{cream}{RGB}{222,217,201}%\usepackage{setspace}

\definecolor{Nathanblue}{rgb}{0.,0.24,0.51}
\newcommand{\blue}{\color{Nathanblue}}

\def\XXint#1#2#3{{\setbox0=\hbox{$#1{#2#3}{\int}$}
		\vcenter{\hbox{$#2#3$}}\kern-.5\wd0}}

\usepackage{setspace}

\def\AmS{{\protect\the\textfont2
        A\kern-.1667em\lower.5ex\hbox{M}\kern-.125emS}}

\makeatletter
\def\thepage{1-\@arabic\c@page}
\def\@pnumwidth{2em}
\makeatother

\newcommand\ba{\begin{eqnarray}}
\newcommand\ea{\end{eqnarray}}
\newcommand\be{\begin{equation}}
\newcommand\ee{\end{equation}}

\newif\ifshowrevisions
\showrevisionstrue
\definecolor{RevisionRed}{rgb}{0.85,0,0}

\begin{document}
\title{{\blue Relativistic bound-state solutions for a non-central Schi{\"o}berg-type hyperbolic potential with double ring-shaped angular terms}}

\author{S.M. Nagiyev{\,}\orcidlink{0000-0003-4214-6721}}
\email{sh.nagiyev@physics.science.az}
\affiliation{\href{https://ror.org/013rnrt24}{Institute of Physics}, Ministry of Science and Education, Javid av. 131, AZ1143, Baku, Azerbaijan}

\author{V. H. Badalov{\,}\orcidlink{0000-0002-5468-1978}}
\email{badalovvatan@yahoo.com}
\affiliation{Department of Theoretical Physics, \href{https://ror.org/054gw3b40}{Baku State University}, Z. Khalilov str. 23, AZ1148, Baku, Azerbaijan}

\noaffiliation

\date{\today}

\begin{abstract}
Singular interactions can reshape quantum spectra by changing admissible wavefunction domains. In this study, we investigate this mechanism in the Klein--Gordon equation for a non-central Schi{\"o}berg-type hyperbolic potential with double ring-shaped angular barriers. Equal scalar and vector couplings separate the radial and angular dynamics. We solve the angular equation exactly on each hemisphere with principal boundary conditions and treat the radial equation with the Greene--Aldrich approximation, obtaining Jacobi-polynomial wavefunctions and an implicit relativistic quantization condition. With this boundary realization, the equatorial inverse-square singularity splits configuration space into reflection-related sectors with identical angular spectra. Reducing the barrier to zero while retaining the equatorial wall selects the odd-reflection spectral branch of the regular full-domain problem; a single hemisphere has no intrinsic parity. The non-relativistic limit recovers the corresponding Schr{\"o}dinger formulation. State-resolved non-relativistic and relativistic benchmarks quantify the centrifugal error. In a dimensionless relativistic example, the relative error in the remaining binding exceeds unity near threshold although the absolute energy error decreases. For NaH and Na$_2$, low-lying central-radial vibrational energies agree reasonably with spectroscopic reference values, whereas the global radial shape limits near-dissociation accuracy and gives an incorrect NaH bound-state count.
\end{abstract}

\maketitle

\section{Introduction}\label{sec1}
\noindent
\lettrine[findent=2pt]{\blue {\textbf{T}}}{}he role that quantum potential models constructed within both non-relativistic and relativistic approaches play in the description of the physical phenomena of the microworld (atoms, molecules, nuclei, and particles), is well known. These models find their wide application in all areas of theoretical physics{\,}\cite{davydov1965, landau1979, flugge1994, greiner2000}. One needs to emphasize that constructing the potential models of quantum physical systems starts with choosing the equation of motion and the subsequent selection of the interaction potential. The Schr{\"o}dinger equation is considered a main equation of motion if one deals with non-relativistic quantum mechanics. Non-relativistic quantum mechanics based on this equation, can explain the main properties of molecular, atomic, and nuclear phenomena in the case of involvement of the various phenomenological interaction potentials. 

At the same time, for a description of the relativistic quantum phenomena, as well as relativistic corrections in nuclear physics and particle physics, one uses the various wave equations of relativistic quantum mechanics. Klein-Gordon (KG), Dirac equations as well as equation of motion of the finite-difference version of the relativistic quantum mechanics belong to such a class of equations{\,}\cite{nagiyev2019}. One needs to highlight here that the relativistic effects also can exhibit themselves in the case when a particle moves in a strong potential field. There are different quantum-mechanical phenomenological interaction potentials. Coulomb interaction, harmonic oscillator, Hulth\'{e}n, Yukawa, Woods-Saxon, Morse, Manning-Rosen, Hartmann, Kratzer, Eckart, Hautot, and Quesne potentials are just a few of them{\,}\cite{nagiyev2024}.

Sometimes a new potential consists of the sum of several known potentials. The ring-shaped potentials belong to such a class. For many years, considerable efforts have been made to obtain the exact or approximate solutions of the non-relativistic, relativistic, and finite-difference equations for various ring-shaped potentials{\,}\cite{quesne1988, berkdemir2006, fu2008, chen2005, mbadjoun2019, berkdemir2009, ahmadov2013, zhang2010, gonul2000, dutra2006, rojas2005, adame1989, zhang2005, lu2005, dong2004, villalba2006, badalov2023, yasuk2006, yi2004, alhaidari2006}. Recent analytic and semi-analytic treatments of Schr\"odinger/Dirac/KG equations with phenomenological potentials continue to provide benchmark bound-state spectra and wavefunctions for molecular and related applications{\,}\cite{Sarathi_2025, Aid_2024, Kadja_2024, de_Oliveira_2021, Boudjedaa_2024, Ahmed_2023}. In the same spirit, earlier relativistic molecular studies often adopt equal scalar--vector coupling together with controlled centrifugal-term approximations{\,}\cite{Maghsoodi_2012, Oluwadare_2012}; following this established line, we analyze a separable double ring-shaped hyperbolic interaction and show that the adopted principal boundary realization at $\vartheta=\pi/2$ yields a hemisphere decomposition. In general, the non-central ring-shaped potentials have the following mathematical expression:
{\fontsize{10pt}{11pt}\selectfont
\be
\label{vrth}
V\left( {r,\vartheta } \right) = V\left( r \right)+\frac{{f\left( \vartheta \right)}}{{r^2 }},
\ee
}
\normalsize
where $V\left( r \right)$ is some central potential, and $f\left( \vartheta \right)$ is some function of the angle characterizing the ring-shaped nature of the potential.

Non-central potentials are good models in quantum chemistry, in nuclear and atomic physics, and in molecular physics. Ring potentials can be used in quantum chemistry to describe organic ring-shaped molecules such as benzene, and in nuclear physics to study the interactions of a deformed pair of nuclei and spin-orbit coupling when particles move in potential fields. This potential is also used as a mathematical model for describing the vibrations of diatomic molecules and is a convenient model in other physical situations. For example, the Hartmann potential{\,}\cite{hartmann1972} is one of the useful combinations in the form of Eq.{\,}\eqref{vrth}, which is used to describe organic molecules in the framework of non-relativistic quantum mechanics. Other important non-central potentials were proposed by Hautot{\,}\cite{hautot1973} to describe the motion of a charged non-relativistic particle in an electric field. The Hartmann potential is a special case of one of the Hautot potentials. In Ref.{\,}\cite{quesne1988}, Quesne investigated a new non-central potential, obtained by replacing the Coulomb part of the Hartmann potential by a harmonic oscillator term.

One of the analytical realizations of such a potential $V\left( r \right)$ is a hyperbolic potential function suggested in Ref.{\,}\cite{schioberg1986}. This potential has a following expression:
{\fontsize{10pt}{11pt}\selectfont
\be
\label{p-hyp}
V\left( r \right) = D\left[ {1 - \sigma _0 \coth \left( {\alpha r} \right)} \right]^2 .
\ee
}
\normalsize
Here, $D$ sets the overall energy scale of the radial well; in the molecular application considered here it is related to the spectroscopic dissociation energy through $D_e = D(1-\sigma_0)^2$. The parameter $\alpha$ is an inverse-range (steepness) parameter, whereas $\sigma_0$ is a dimensionless shape parameter controlling the position and shape of the minimum. The
potential has its minimum value $V(r_e)=0$ at
\[
r_e=\frac{1}{\alpha}\operatorname{artanh}\sigma_0,
\]
and approaches $D_e$ exponentially as $r\to\infty$. Its various properties have already been studied in Refs.{\,}\cite{jun2005,dong2007,ikhdair2009}.

At the same time, exact solutions of the Schr{\"o}dinger equation have been discussed for the double ring-shaped harmonic oscillator potential in Refs.{\,}\cite{caprio1989,aktas2009,chen2013},
{\fontsize{10pt}{11pt}\selectfont
\be
\label{p-drsho}
V \left( {r,\vartheta } \right) = \frac{1}{2}kr^2  + \frac{1}{{r^2 }}\left( {\frac{\beta }{{\sin ^2 \vartheta }} + \frac{\nu }{{\cos ^2 \vartheta }}} \right),
\ee
}
\normalsize
and Yasuk \textit{et al.}{\,}\cite{yasuk2008} presented an exact KG solution for the same type of ring-shaped oscillator. Maghsoodi \textit{et al.}{\,}\cite{maghsoodi2013} studied the Dirac equation with a P{\"o}schl--Teller double ring-shaped Coulomb potential, while Durmus \textit{et al.}{\,}\cite{durmus2007} considered relativistic and non-relativistic solutions for two-atom molecules in a double-ring Kratzer potential. Among these studies, Chen \textit{et al.}{\,}\cite{chen2013} emphasized a limiting-case consistency problem in the angular sector of the double ring-shaped harmonic oscillator in Eq.{\,}\eqref{p-drsho}. The difficulty is that some double-ring solutions do not reduce to the single ring-shaped oscillator when $\nu=0$, or to the ordinary harmonic oscillator when both $\nu=0$ and $\beta=0$. Their analysis traced this problem to the angular wavefunctions and proposed super-universal associated Legendre polynomials as a way to recover the usual limiting cases.

\begin{figure}
\centering
\includegraphics[width=\linewidth]{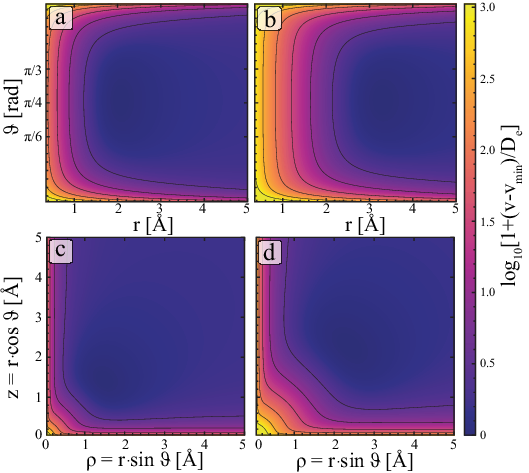}
\caption{\textbf{Potential maps in spherical and cylindrical coordinates.} Panels \textbf{(a,c)} and \textbf{(b,d)} use the NaH and Na$_2$ radial parametrisations,respectively. The upper row shows $V(r,\vartheta)$ for $0<\vartheta<\pi/2$, and the lower row shows $V(\rho,z)$ for $\rho,z>0$, where $\rho=r\sin\vartheta$ and $z=r\cos\vartheta$. The two rows use different spatial plotting windows. The color scale represents $\log_{10}[1+(V-V_{\min})/D_e]$, with $D_e$ the corresponding radial dissociation energy and
$V_{\min}$ the reference energy subtracted from the potential. The equatorial singularity occurs at $z=0$ for $\nu+\nu'>0$;
the polar-axis singularity occurs at $\rho=0$ when $\beta+\beta'>0$. Only the $z>0$ hemisphere is shown.}\label{fig1}
\end{figure}

Motivated by this angular-domain issue, and by the Schi\"oberg-type radial interaction used in molecular applications, we consider the following non-central potential:
{\fontsize{8.3pt}{8pt}\selectfont
\be
\label{new-drsho}
\begin{split}
V\left({\vec r} \right)\equiv V \left( {r,\vartheta } \right) = & D\left[ {1 - \sigma _0 \coth \left( {\alpha r} \right)} \right]^2  + \frac{1}{{r^2 }}\left( {\frac{\beta+\beta'\cos^2 \vartheta }{{\sin ^2 \vartheta }} + \frac{\nu+\nu'\sin^2 \vartheta }{{\cos ^2 \vartheta }}} \right),\\
\beta,\;\beta',\;\nu,\;\nu'&\geq0;\quad\nu+\nu'>0.
\end{split}
\ee
}
\normalsize
\noindent
The radial component of Eq.{\,}\eqref{new-drsho} belongs to the Schi\"oberg-type hyperbolic class, whereas the angular component contains two inverse-square barriers. Its $\sin^{-2}\vartheta$ contribution is connected with the usual ring-shaped term, while the additional $\cos^{-2}\vartheta$ contribution produces an interior singularity at $\vartheta=\pi/2$. The radial part is therefore related to the hyperbolic model considered in Ref.{\,}\cite{nagiyev2024}, but the present work addresses a different angular problem generated by the double-ring sector structure. For $\nu+\nu'>0$, under the principal separated boundary conditions adopted below, the singularity at $\vartheta=\pi/2$ separates the full interval $0<\vartheta<\pi$ into two disconnected sectors, $0<\vartheta<\pi/2$ and $\pi/2<\vartheta<\pi$. Since these sectors are mapped onto one another by the reflection $\vartheta\rightarrow\pi-\vartheta$, it is sufficient to solve the angular equation in the first sector; the second sector gives mirror-related states with the same separation eigenvalues. Similar domain separation is also encountered in singular oscillator systems{\,}\cite{calogero1969}.

The analytic methods used here are established. In particular, Ref.{\,}\cite{nagiyev2024} already combines the same hyperbolic radial interaction with a different ring-shaped angular term under equal scalar--vector KG coupling, using a centrifugal approximation and Jacobi polynomials. Double-ring oscillator, Kratzer, and relativistic systems were studied in Refs.{\,}\cite{caprio1989, aktas2009, chen2013,yasuk2008,durmus2007,maghsoodi2013}; parity properties were also examined in Ref.{\,}\cite{sun2015parity}. Singular-boundary quantization and domain-sensitive zero-coupling limits are established principles{\,}\cite{gitman2010, tsutsui2003, feher2005}. The present contribution is their explicit application to this radial--angular combination, together with state-resolved validation of the centrifugal replacement. Neither the Jacobi method, the general mechanism of sector separation, nor the Lambert-$W$ parametrization{\,}\cite{wang2012} is claimed as new.

The aim of this work is therefore threefold. First, we derive the separated KG equations for the non-central Schi\"oberg-type hyperbolic potential in Eq.{\,}\eqref{new-drsho}. Second, we obtain the regular angular solutions in the half-domain, clarify the mirror-sector interpretation, and discuss the zero-barrier limiting case. Third, we derive the corresponding radial wavefunctions, energy equation, non-relativistic limit, and vibrational spectra for NaH and Na$_2$. This paper is organized as follows. In Section{\,}\ref{sec2}, we solve the radial and angular parts of the KG equation for the non-central potential in Eq.{\,}\eqref{new-drsho} and obtain the relativistic energy equation and wavefunctions. Section{\,}\ref{sec3} gives the non-relativistic limit. Section{\,}\ref{sec4} is devoted to numerical results and discussion for diatomic molecules. Finally, we summarize the main conclusions in Section{\,}\ref{sec5}.

\section{The Solution of Klein-Gordon equation for non-central potential}\label{sec2}
\subsection{Separation of the variables of the Klein-Gordon equation}
\noindent
Our starting point is the KG equation with scalar $V_1\left(\vec r \right)$ and vector $V_2\left(\vec r \right)$ potentials:
{\fontsize{10pt}{11pt}\selectfont
\be
\label{kg-1}
\left\{ { - \hbar ^2 c^2 \nabla ^2  + \left[ {Mc^{2} + V_1 \left( {\vec r} \right)} \right]^2  - \left[ {E - V_2
\left( {\vec r} \right)} \right]^2 } \right\}\psi \left( {\vec r} \right) = 0,
\ee
}
\normalsize
where $M$ and $E$ are the mass and energy of the relativistic particle. For simplicity, we adopt the equal scalar--vector coupling
\[
V_1(\vec r)=V_2(\vec r)\equiv \frac{1}{2}V(\vec r).
\]
This choice is used here as a standard solvable relativistic limit rather than as a unique microscopic prescription. It yields a tractable KG problem with a well-defined non-relativistic reduction and, in the present non-central setting, preserves the separable structure needed to isolate the physical effect of the double ring-shaped angular barrier, in particular the sector splitting induced by the singular point at $\vartheta=\pi/2$. At the non-relativistic level, writing
\[
E = Mc^2 + E_N, \qquad E - Mc^2 \to E_N, \qquad E + Mc^2 \to 2M c^2,
\]
Eq.{\,}\eqref{kg-1} reduces to the corresponding Schr\"odinger equation for the potential \(V(\vec r)\); the detailed \(c\to\infty\) reduction is presented in Sec.{\,}\ref{sec3}. For the molecular non-relativistic illustration, the one-body mass parameter is then identified with the reduced mass $M=\mu$. A treatment with unequal scalar and vector couplings would generally require a different, typically numerical or perturbative, framework and lies beyond the scope of the present analytic study.

Now the introduction of the notations
$
\varepsilon ^2  = \frac{{E^2  - M^2 c^4 }}{{\hbar ^2 c^2 }}, \\
\gamma  = \frac{{{E + Mc^2 }}}{{\hbar ^2 c^2 }},
$
and substitution of Eq.{\,}\eqref{new-drsho} into Eq.{\,}\eqref{kg-1} slightly changes it as follows:
{\fontsize{10pt}{9pt}\selectfont
\be
\label{kg-2}
\left[ {\nabla ^2  + \varepsilon ^2  - \gamma V\left( r \right) - \gamma \frac{{f\left( \vartheta  \right)}}{{r^2 }}} \right]\psi \left( {\vec r} \right) = 0,
\ee
}
\normalsize
with
{\fontsize{9pt}{9pt}\selectfont
\be
\label{nabla-2}
\nabla ^2  = \frac{1}{{r^2 }}\partial _r \left( {r^2 \partial _r } \right) + \frac{{\Delta _{\vartheta ,\varphi } }}{{r^2 }},\quad \Delta _{\vartheta ,\varphi }  = \frac{1}{{\sin \vartheta }}\partial _\vartheta  \left( {\sin \vartheta \partial _\vartheta  } \right) + \frac{1}{{\sin ^2 \vartheta }}\partial _\varphi  ^2 .
\ee
}
\normalsize
Because Eq.{\,}\eqref{kg-2} contains the non-central interaction in the separable form $V(r,\vartheta)=V(r)+f(\vartheta)/r^{2}$, the Klein--Gordon equation separates exactly in spherical coordinates. Therefore, the wavefunction can
be written in the factorized form
{\fontsize{10pt}{9pt}\selectfont
\be
\label{wf-1}
\psi \left( {\vec r} \right) = \frac{{\chi \left( r \right)}}{r}\Theta \left( \vartheta  \right)\Phi \left( \varphi  \right).
\ee
}
\normalsize
It should be noted that since the operator $\hat L_\varphi   =  - i\hbar \partial _\varphi$ commutes with the stationary differential expression in Eq.{\,}\eqref{kg-2}, the function $\Phi \left( \varphi \right)$ has the following standard form:
{\fontsize{10pt}{9pt}\selectfont
\be
\label{phi-1}
\begin{split}
 \Phi \left( \varphi  \right)  \equiv \Phi _m \left( \varphi  \right) = \frac{1}{{\sqrt {2\pi } }}e^{im\varphi } ,  \quad  m = 0, \pm 1, \pm 2, \ldots .
\end{split}
\ee
}
\normalsize
After substituting Eq.{\,}\eqref{wf-1} into Eq.{\,}\eqref{kg-2}, the following system of second-order differential equations is obtained:
{\fontsize{10pt}{9pt}\selectfont
\be
\label{kg-3}
\chi ''\left( r \right) + \left[ {\varepsilon ^2  - \gamma V\left( r \right) - \frac{\Lambda }{{r^2 }}} \right]\chi \left( r \right) = 0,
\ee
}
{\fontsize{7.8pt}{9pt}\selectfont
\begin{align}
& \Theta ''\left( \vartheta  \right) + \cot \vartheta  \Theta '\left( \vartheta  \right) + \left[ {\Lambda  - \frac{{m^2  + \gamma \left(\beta+\beta'\cos^2 \vartheta\right) }}{{\sin ^2 \vartheta }} - \frac{{\gamma \left(\nu+\nu'\sin^2 \vartheta\right) }}{{\cos ^2 \vartheta }}} \right]\Theta \left( \vartheta  \right) = 0, \label{kg-4}
\end{align}
}
\normalsize
where $\Lambda$ is an another separation constant. Let's define the generalized orbital quantum number $L$, which in the general case is not an integer: $\Lambda=L\left(L+1\right)$. Eq.{\,}\eqref{kg-3} yields the following effective potential:
{\fontsize{10pt}{9pt}\selectfont
\be
\label{v-eff}
V_{eff} \left( r \right) = \gamma D\left[ {1 - \sigma_{0} \coth \left( {\alpha r} \right)} \right]^2  + \frac{\Lambda }{{r^2 }}.
\ee
}
\normalsize
In the next part, we will look for analytical bound-state solutions $\chi \left( r \right)$ of the radial part of the KG Eq.{\,}\eqref{kg-3}.

\subsection{The Radial Klein-Gordon equation}
\noindent
Let us rewrite the radial differential Eq.{\,}\eqref{kg-3} in expanded form as:
{\fontsize{10pt}{9pt}\selectfont
\be
\label{kg-5}
\chi ''\left( r \right) + \left\{ {\varepsilon ^2  - \gamma D\left[ {1 - \sigma _0 \coth \left( {\alpha r} \right)} \right]^2  - \frac{\Lambda }{{r^2 }}} \right\}\chi \left( r \right) = 0.
\ee
}
\normalsize
For \(\Lambda \neq 0\), Eq.{\,}\eqref{kg-5} is not exactly solvable for the present
hyperbolic potential in closed form. To that,  one of the frequently used approximate methods for solving equations for any $l \neq 0$ value of orbital quantum number of motion with a central or non-central potential are the Greene-Aldrich{\,}\cite{greene1976} and Pekeris approximations {\,}\cite{pekeris1934}. In order to obtain an analytical solution to this equation, we apply the Greene-Aldrich approximation to the centrifugal potential{\,}\cite{greene1976}:
{\fontsize{10pt}{9pt}\selectfont
\be 
\label{kg-6}
\frac{1}{{r^2 }} \approx \frac{{4\alpha ^2 e^{ - 2\alpha r} }}{{\left( {1 - e^{ - 2\alpha r} } \right)^2 }} ,  \quad  \alpha r<<1.
\ee
}
\normalsize
Its substitution into Eq.{\,}\eqref{kg-5} yields:
{\fontsize{10pt}{9pt}\selectfont
\be
\label{kg-51}
\chi ''\left( r \right) + \left[ {\varepsilon ^2  - \gamma D\left( {1 - \sigma _0 \frac{{1 + e^{ - 2\alpha r} }}{{1 - e^{ - 2\alpha r} }}} \right)^2 - \frac{4\alpha^2 \Lambda e^{-2\alpha r}}{(1-e^{-2\alpha r})^2}} \right]\chi \left( r \right) = 0.
\ee
}
\normalsize
Now, due to a function $e^{ - 2\alpha r}$ appears in the equation, it is convenient to introduce a new variable $z = e^{ -2\alpha r}$. Then, we can represent Eq.{\,}\eqref{kg-51} in the following form:
{\fontsize{10pt}{9pt}\selectfont
\be
\label{kg-7}
\sigma \left( z \right)\chi ''\left( z \right) +
\tilde \tau \left( z \right)\chi '\left( z \right) + \frac{{\tilde
\sigma \left( z \right)}}{{\sigma \left( z \right)}}\chi \left( z
\right) = 0,
\ee
}
\normalsize
where $\sigma \left( z \right) = z\left( {1 - z}
\right)$, $\tilde \tau \left( z \right) = 1 - z$ and $\tilde \sigma
\left( z \right) = \left( {\frac{1}{4} - c_2 } \right)z^2  + c_1 z -
c_0$. Here, the coefficients $c_i$ ($i=0,1,2$) are defined through the following expressions:
{\fontsize{10pt}{9pt}\selectfont
\be
\label{kg-8}
\begin{split}
 c_2  = \frac{1}{4} - \frac{{\varepsilon ^2 }}{{4\alpha ^2 }} + \frac{{\gamma D}}{{4\alpha ^2 }}\left( {1 + \sigma _0 } \right)^2 , \\
 c_1  =  - \frac{{\varepsilon ^2 }}{{2\alpha ^2 }} + \frac{{\gamma D}}{{2\alpha ^2 }}\left( {1 - \sigma _0 ^2 } \right) - \Lambda , \\
 c_0  =  - \frac{{\varepsilon ^2 }}{{4\alpha ^2 }} + \frac{{\gamma D}}{{4\alpha ^2 }}\left( {1 - \sigma _0 } \right)^2 .
\end{split}
\ee
}
\normalsize
We look for solution of Eq.{\,}\eqref{kg-7} in the form
{\fontsize{10pt}{9pt}\selectfont
\be
\label{kg-9}
\chi \left( z \right) = \varphi \left( z \right)y\left( z \right),\quad \varphi\left( z \right) = z^A \left( {1 - z} \right)^B ,
\ee
}
\normalsize
where $A$ and $B$ are unknown constant parameters. From the boundary conditions $\chi \left( {z = 0} \right) = \chi \left( {z = 1} \right) = 0$ it follows that $A>0$ and $B>0$ to be hold. Substitution of Eq.{\,}\eqref{kg-9} at Eq.{\,}\eqref{kg-7} yields the following equation for the function $y\left( z \right)$:
{\fontsize{8.8pt}{9pt}\selectfont
\be
\label{kg-91}
z\left( {1 - z} \right)y''\left( z \right) + \left[ {2A + 1 - \left( {2A + 2B + 1} \right)z} \right]y'\left( z \right) + \frac{{\gamma _2 z^2  + \gamma _1 z + \gamma _0 }}{{z\left( {1 - z} \right)}}y\left( z \right) = 0.
\ee
}
\normalsize
Here, the following notations are introduced for $\gamma_i$ ($i=0,1,2$) coefficients:
{\fontsize{9pt}{9pt}\selectfont
\be
\label{kg-92}
\begin{split}
 \gamma _0  &= A^2  - c_0 , \quad   \gamma _1  = c_1  - B - 2A\left( {A + B} \right), \quad     \gamma _2  = \left( {A + B} \right)^2  - c_2  + \frac{1}{4}.
\end{split}
\ee
}
\normalsize
In order to obtain polynomial solutions for function $y\left( z \right)$, one needs to get the additional condition
{\fontsize{10pt}{9pt}\selectfont
\be
\label{kg-93}
\frac{{\gamma _2 z^2  + \gamma _1 z + \gamma _0 }}{{z\left( {1 - z} \right)}} = \lambda=const ,
\ee
}
\normalsize
which satisfies the following conditions:
{\fontsize{10pt}{9pt}\selectfont
\be
\label{kg-94}
\begin{split}
 \gamma _0  = 0 , \quad   \gamma _1 + \gamma _2 = 0, 
 \end{split}
\ee
}
\normalsize
Then, from above conditions one obtains that
{\fontsize{10pt}{9pt}\selectfont
\be
\label{kg-10}
A = \sqrt {c_0 } ,\quad B = \frac{1}{2} + \sqrt {c_0 + c_2 - c_1 } ,
\ee
}
\normalsize
with $c_0  + c_2  - c_1  = \gamma D\sigma _0 ^2 /\alpha ^2  + \Lambda  + 1/4 > 0$. The condition $c_0>0$ (i.e., $A>0$) imposes additional restriction to the upper values of the energy spectrum as follows:
\[
E < Mc^2  + D\left( {1 - \sigma _0 } \right)^2,
\]
wherein the function $y\left( z \right)$ now satisfies the following equation:
{\fontsize{10pt}{9pt}\selectfont
\be
\label{kg-11}
z\left( {1 - z} \right)y''\left( z \right) + \left[ {2A + 1 - \left( {2A + 2B + 1} \right)z} \right]y'\left( z \right) + \lambda y\left( z \right) = 0.
\ee
}
\normalsize
In here, the free parameter $\lambda$ is:
{\fontsize{10pt}{9pt}\selectfont
\be
\label{kg-12}
\lambda  = c_1  - B - 2A\left( {A + B} \right).
\ee
}
\normalsize
Let us compare Eq.{\,}\eqref{kg-11} with the following equation for the Jacobi polynomials $\bar y\left( x \right) = P_n^{\left( {\bar \alpha ,\bar \beta } \right)} \left( x \right)$ {\,}\cite{koekoek2010}:
{\fontsize{8.4pt}{9pt}\selectfont
\be
\label{kg-13}
\begin{split}
\left( {1 - x^2 } \right)\bar y''\left( x \right) + \left[ {\bar \beta  - \bar \alpha  - \left( {\bar \alpha  + \bar \beta  + 2} \right)x} \right]\bar y'\left( x \right)+ n\left( {n + \bar \alpha  + \bar \beta  + 1} \right)\bar y\left( x \right) = 0.
\end{split}
\ee
}
\normalsize
As a result of this comparison, one observes that
{\fontsize{8.6pt}{9pt}\selectfont
\be
\label{kg-14}
\begin{split}
\bar \alpha  = 2A, \quad   \bar \beta  = 2B - 1,  \quad    \lambda  = n\left( {n + \bar \alpha  + \bar \beta  + 1} \right),  \quad  n = 0,1,2, \ldots, n_{\max},
\end{split}
\ee
}
\normalsize
as well as
{\fontsize{10pt}{9pt}\selectfont
\be
\label{kg-15}
y\left( z \right) \equiv y_n \left( z \right) = P_n^{\left( {2A,2B - 1} \right)} \left( {1 - 2z} \right),
\ee
}
\normalsize
with
{\fontsize{10pt}{9pt}\selectfont
\be
\label{nmax}
n\in\mathbb{N}_0,\qquad n < \frac{\sqrt{\gamma(E)D\sigma_0}}{\alpha}-B(E).
\ee
}
\normalsize
This inequality is strict and must be checked at each candidate KG energy; its right-hand side is energy dependent and is not itself an integer quantum number. On the chosen branch, $-Mc^2<E<Mc^2+D(1-\sigma_0)^2$ and the unsquared decay parameter must remain positive.

One equates expressions for $\lambda$ from Eq.{\,}\eqref{kg-12} and Eq.{\,}\eqref{kg-14} and obtains an analytical expression for the energy eigenvalues in the form:
{\fontsize{8.3pt}{9pt}\selectfont
\be
\label{e-sp}
E^2  - M^2 c^4  = \hbar ^2 c^2 \gamma D\left( {1 - \sigma _0 } \right)^2 - \hbar^2c^2 \alpha ^2 \left[ {\frac{p\sigma _0 \left( {1 - \sigma _0 } \right) - \Lambda  - n^2  - \left( 2n + 1 \right)B}{ {n + B} }} \right]^2 ,
\ee
}
\normalsize
where $p=\gamma D/\alpha^2$, and the parameter $\Lambda$ is going to be determined by the equation of the angular part of the KG equation. Substitution of $\varphi \left( z \right)$ and $y_n \left( z \right)$ into Eq.{\,}\eqref{kg-9} for the radial wavefunction gives the expression as:
\begin{equation}
\begin{split}
\chi_n(z)&=C_n z^A(1-z)^B P_n^{(2A,2B-1)}(1-2z),\\
C_n^2&=\frac{4\alpha A\,n!\,(n+A+B)}{n+B}
\cdot\frac{\Gamma(n+2A+2B)}{\Gamma(n+2B)\Gamma(n+2A+1)}.
\end{split}
\label{wf-theta}
\end{equation}
Here, $A=\sqrt{c_0}$ and $B=1/2+\sqrt{c_0+c_2-c_1}$. The normalization is
\begin{equation}
\int_0^\infty|\chi_n(r)|^2\,dr=\frac{1}{2\alpha}\int_0^1|\chi_n(z)|^2\frac{dz}{z}=1.
\end{equation}
This is a convenient $L^2$ normalization of a separated factor. It is not the KG charge normalization. For exact stationary KG solutions with distinct energies, subtraction of the KG equations, with vanishing boundary terms, gives
\begin{equation}
\int [E_i+E_j-V(\mathbf r)]\psi_i^*(\mathbf r)\psi_j(\mathbf r)\,d^3r=0\quad (E_i\ne E_j).
\label{eq:KG_weight_revision}
\end{equation}

The radial normalization integral is evaluated by using the following known integral{\,}\cite{abramowitz1964}:
{\fontsize{8.5pt}{9pt}\selectfont
\be
\label{kls-01}
\begin{split}
\int_{0}^{1} (1 - z)^{2(\delta + 1)} z^{2\lambda - 1}\, & \biggl[ {}_{2}F_{1}\!\bigl({\begin{array}{*{20}c}
   { - n,2\left( {\delta  + \lambda  + 1} \right) + n}, {2\lambda  + 1}, 
\end{array};z} \bigr) \biggr]^2 dz \\
&\quad = \frac{(n + \delta + 1)\,n!\,\Gamma(n + 2\delta + 2)\,\Gamma(2\lambda)\,\Gamma(2\lambda + 1)}
     {(n + \delta + \lambda + 1)\,\Gamma(n + 2\lambda + 1)\,\Gamma(2(\delta + \lambda + 1) + n)}.
\end{split}
\ee
}
\normalsize
with $\;  \delta  >  - \frac{3}{2},\quad \lambda  > 0 .$

\subsection{The angular part of the Klein-Gordon equation}
\noindent
Now, let us find the solution to the angular part of the KG Eq.{\,}\eqref{kg-4}. As a result of substitution $\Theta \left( \vartheta  \right) = \frac{1}{{\sqrt {\sin \vartheta } }}\Theta _1 \left( \vartheta  \right)$, its first order derivative disappears and the equation simplifies as follows:
{\fontsize{10pt}{9pt}\selectfont
\be
\label{kgt-1}
\Theta _1 ''\left( \vartheta  \right) + \left( {\Lambda _1  - \frac{{\rho _1 }}{{\sin ^2 \vartheta }} - \frac{{\rho _2 }}{{\cos ^2 \vartheta }}} \right)\Theta _1 \left( \vartheta  \right) = 0,
\ee
}
\normalsize
where $\Lambda _1  = \Lambda  + 1/4+\gamma \left( {\beta ' + \nu '} \right)$, $\rho _1  = m^2  + \gamma \left(\beta+\beta'\right)  - 1/4$ and $\rho _2  = \gamma \left(\nu+\nu'\right) $.

To clarify the domain-dependent limiting behavior, we start with the case, where $\nu=\nu'=0$. Then, one obtains
{\fontsize{10pt}{9pt}\selectfont
\be
\label{kgh-1}
{\rm H}_1 '' \left( \vartheta  \right) + \left( {\tilde \Lambda _1  - \frac{{\rho _1 }}{{\sin ^2 \vartheta }}} \right){\rm H}_1 \left( \vartheta  \right) = 0,\quad 0 \le \vartheta  \le \pi ,
\ee
}
\normalsize
where $\tilde \Lambda _1  = \Lambda  + \frac{1}{4} + \gamma \beta '$ and $\Theta _1 \left( \vartheta  \right)\mathop  \to \limits^{\nu  = \nu ' = 0} {\rm H}_1 \left( \vartheta  \right)$. Next, introducing a new variable $x=\cos^2\frac{\vartheta}{2}$ in Eq.{\,}\eqref{kgh-1} one obtains
{\fontsize{10pt}{9pt}\selectfont
\be
\label{kgh-2}
x\left( {1 - x} \right){\rm H}_1 '' \left( x \right) + \left( \frac{1}{2} - x \right){\rm H}_1 ' \left( x \right) + \left[ {\tilde \Lambda _1  - \frac{{\rho _1 }}{{4x\left( {1 - x} \right)}}} \right]{\rm H}_1 \left( x \right) = 0.
\ee
}
\normalsize
We look for exact restricted solution of Eq.{\,}\eqref{kgh-2} as follows:
{\fontsize{10pt}{9pt}\selectfont
\be
\label{kgh-3}
{\rm H}_1 \left( x \right) = x^{A_1 } \left( {1 - x} \right)^{A_1 } {\rm H}_2 \left( x \right).
\ee
}
\normalsize
Further, choosing $A_1  = \frac{1}{4} + \frac{1}{2}\sqrt {m^2  + \gamma \left(\beta +\beta'\right) }$ yields
{\fontsize{8.8pt}{9pt}\selectfont
\be
\label{kgh-4}
x\left( {1 - x} \right){\rm H}_2 ''\left( x \right) + \left[ {2A_1  + \frac{1}{2} - \left( {4A_1  + 1} \right)x} \right]{\rm H}_2 '\left( x \right) + \left( {\tilde \Lambda _1  - 4A_1 ^2 } \right){\rm H}_2 \left( x \right) = 0.
\ee
}
\normalsize
Introduction of a new variable $t=2x-1$ changes Eq.{\,}\eqref{kgh-4} as follows:
{\fontsize{10pt}{9pt}\selectfont
\be
\label{kgh-5}
\left( {1 - t^2 } \right){\rm H}_2 ''\left( t \right) - \left( {4A_1  + 1} \right)t{\rm H}_2' \left( t \right) + \left( {\tilde \Lambda _1  - 4A_1 ^2 } \right){\rm H}_2 \left( t \right) = 0.
\ee
}
\normalsize
Its comparison with Eq.{\,}\eqref{kg-13} allows us to write down the solution ${\rm H}_2 \left( t \right)$ in terms of the Jacobi polynomials as follows:
{\fontsize{10pt}{9pt}\selectfont
\be
\label{kgh-6}
{\rm H}_2 \left( t \right) = P_k ^{\left( {\tilde \alpha ,\tilde \alpha } \right)} \left( t \right),
\ee
}
\normalsize
with
{\fontsize{10pt}{9pt}\selectfont
\be
\label{kgh-7}
\tilde \alpha  = 2A_1  - \frac{1}{2},\quad \tilde \Lambda _1  = \left( k + 2A_1 \right)^2 ,\quad k = 0,1,2, \ldots .
\ee
}
\normalsize
As a result, one can obtain that for the function ${\rm H}\left( \vartheta  \right) = \frac{1}{{\sqrt {\sin \vartheta } }}{\rm H}_1 \left( \vartheta  \right)$
{\fontsize{10pt}{9pt}\selectfont
\be
\label{kgh-8}
{\rm H}_k \left( \vartheta  \right)\equiv {\rm H}\left( \vartheta  \right) =\tilde N_k  \left( {\sin \vartheta } \right)^{2A_1  - \frac{1}{2}} P_k ^{\left( {2A_1  - \frac{1}{2},2A_1  - \frac{1}{2}} \right)} \left( {\cos \vartheta } \right).
\ee
}
\normalsize
At fixed $E$, the normalization and orthogonality of these Jacobi functions are specified by
{\fontsize{10pt}{9pt}\selectfont
\[
\int\limits_0^{\pi} { \rm H_k \left( \vartheta  \right) \rm H_{k'} \left( \vartheta  \right)\sin \vartheta d\vartheta }  = \delta _{k,k'}, 
\]
}
\normalsize
one obtains that
{\fontsize{10pt}{9pt}\selectfont
\be
\label{kgt-81}
\tilde N_k  =\frac{2^{-2A_1+\frac{1}{2}} \sqrt {\left(k +2A_1\right)\Gamma \left( k + 4A_1 \right)k!}}{\Gamma \left( k + 2A_1  + \frac{1}{2}
\right)}.
 \ee
}
\normalsize
Then, one can also observe that
{\fontsize{10pt}{9pt}\selectfont
\be
\label{kgh-9}
\Lambda  \equiv \Lambda_k = \left( {k + 2A_1 } \right)^2  - \frac{1}{4} - \gamma \beta ' = L_k \left( {L_k  + 1} \right) , 
\ee
}
\normalsize
where the generalized orbital momentum $L_k$ equals to
{\fontsize{10pt}{9pt}\selectfont
\be
\label{kgh-96}
L_k  =\sqrt{\left( {k +2 A_1  } \right)^2 - \gamma \beta '} -\frac{1}{2}.
\ee
}
\normalsize 
If we consider the following case $\beta=\beta'=0$, one will obtain $L_k  = l = k + \left| m \right|$, which means that all possible values of the orbital momentum will be realized. Also, one need to write down the equation for the energy spectrum that corresponds to the solution Eq.{\,}\eqref{kgh-8}. It is as follows:
\footnotesize
\be
\begin{split}
E^2 - M^2 c^4 &= \hbar^2 c^2 \gamma D\,(1 - \sigma_0)^2 \\
&\quad - \hbar^2 c^2 \alpha^2 \left[
    \frac{
      p\,\sigma_0\,(1 - \sigma_0) - (k + 2A_1)^2 + \gamma \beta' + \tfrac{1}{4}
      - n^2 - (2n + 1) B
    }{n + B}
  \right]^2.
\end{split}
\label{kgh-10}
\ee
\normalsize
\subsection{Angular domain and equatorial boundary conditions}
\noindent
The following endpoint analysis is performed at fixed real $E>-Mc^2$, so that $\gamma(E)>0$. Set

{\fontsize{9pt}{9pt}\selectfont
\begin{equation}
a=\sqrt{m^2+\gamma(\beta+\beta')},\quad b=\sqrt{\tfrac14+g},\quad g=\gamma(\nu+\nu'),\quad u=\sqrt{\sin\vartheta}\,\Theta.
\label{eq:ab_revision}
\end{equation}
}
\normalsize 
Near the equator, $y=|\vartheta-\pi/2|$, the two local solutions are
\begin{equation}
u\sim C_+y^{1/2+b}+C_-y^{1/2-b}.
\label{eq:endpoint_revision}
\end{equation}
For $g\geq3/4$, the endpoint is limit-point and only the principal branch is locally square integrable. For $0<g<3/4$, both branches are locally square integrable and a self-adjoint boundary condition must be specified{\,}\cite{gitman2010,gesztesy2024}. We choose $C_-=0$ independently on both sides. This is the principal (Friedrichs) realization, selected by finite local quadratic-form energy for the repulsive equatorial term. With reflected pole conditions, the angular operator is the direct sum of two identical hemisphere operators and carries no current across the equator. Sector disconnection is therefore a statement about this realization, not about every extension of the differential expression.

In the limit-circle regime $0<g<3/4$, other angular self-adjoint extensions can couple the two boundary channels and change the spectrum{\,}\cite{tsutsui2003}. Their compatibility with the full KG charge space and with a microscopic regularization requires further analysis and is not asserted here. At the pole we select $u\sim\vartheta^{1/2+a}$; for $a=0$ the non-logarithmic $\sqrt{\vartheta}$ branch is retained. The critical case $A_1=1/4$ is therefore allowed. Fixed-energy angular orthogonality must not be confused with orthogonality between different physical KG energies. The angular eigenvalue branches and representative principal densities are illustrated in Supplementary (SM) Fig.{\,}S2.

Now, one can return back to Eq.{\,}\eqref{kgt-1}. The term $\frac{\rho_2}{\cos^2 \vartheta}$ in Eq.{\,}\eqref{kgt-1} turns into infinity when $\vartheta=\pi/2$ (see Fig.{\,}\ref{fig1}). So we consider Eq.{\,}\eqref{kgt-1} only in the interval $0<\vartheta < \pi/2$ with the boundary conditions $\Theta_1 \left(0\right)=\Theta_1 \left(\pi/2\right)=0$. To solve it, one introduces a new variable $x=\sin^2\vartheta$. Then, one can observe that
{\fontsize{9pt}{9pt}\selectfont
\be
\label{kgt-2}
x\left( {1 - x} \right)\Theta _1 ''\left( x \right) + \left( {\frac{1}{2} - x} \right)\Theta _1 '\left( x \right) + \frac{1}{4}\left( {\Lambda _1  - \frac{{\rho _1 }}{x} - \frac{{\rho _2 }}{{1 - x}}} \right)\Theta _1 \left( x \right) = 0.
\ee
}
\normalsize
We look for the following solution of Eq.{\,}\eqref{kgt-2}:
{\fontsize{10pt}{9pt}\selectfont
\be
\label{kgt-3}
\Theta _1 \left( x \right) = x^{A_1 } \left( {1 - x} \right)^{B_1 } \Theta _2 \left( x \right),
\ee
}
\normalsize
where
{\fontsize{10pt}{9pt}\selectfont
\be
\label{kgt-4}
A_1  = \frac{1}{4} + \frac{1}{2}\sqrt {m^2  + \gamma \left(\beta+\beta'\right)} , \quad  B_1  = \frac{1}{4} + \frac{1}{2}\sqrt {\frac{1}{4}  + \gamma \left(\nu+\nu'\right) } .
\ee
}
\normalsize
The selected principal branches have $A_1\geq1/4$ and $B_1>1/2$ for the repulsive equatorial barrier. At the critical pole $A_1=1/4$, the non-logarithmic branch is used. 
In terms of a new variable $t=2x-1=2\sin^2 \vartheta-1$, the function $\Theta_2 \left(x\right)$ obeys the following equation:
{\fontsize{9pt}{9pt}\selectfont
\be
\label{kgt-5}
\left( {1 - t^2 } \right)\Theta _2 ''\left( t \right) + \left[ {2A_1  - 2B_1  - \left( {2A_1  + 2B_1  + 1} \right)t} \right]\Theta _2 '\left( t \right) + \mu \Theta _2 \left( t \right) = 0,
\ee
}
\normalsize
where
{\fontsize{10pt}{9pt}\selectfont
\be
\label{kgt-6}
\mu  = \frac{1}{4}\left[\Lambda+\gamma\left(\beta'+\nu'\right)\right]  + \frac{1}{{16}} - \left( {A_1  + B_1 } \right)^2 .
\ee
}
\normalsize
While we compare Eq.{\,}\eqref{kgt-5} with Eq.{\,}\eqref{kg-13}, we yield:
{\fontsize{8.2pt}{8pt}\selectfont
\be
\label{kgt-7}
\begin{split}
\Theta _2 \left( x \right) = P_k^{\left( {2A_1  - 1/2,2B_1  - 1/2} \right)} \left( {1 - 2x} \right),\quad \mu  = k\left( {k + 2A_1  + 2B_1 } \right),\quad k = 0,1,2, \ldots .
\end{split}
\ee
}
\normalsize
This means that the angular wavefunction is also expressed through the Jacobi polynomials as follows:
\footnotesize
\be
\label{kgt-71}
\Theta _k \left( \vartheta  \right) \equiv \Theta \left( \vartheta  \right)= N_k \left( {\sin \vartheta } \right)^{2A_1  - \frac{1}{2}} \left( {\cos \vartheta } \right)^{2B_1 } P_k ^{\left( {2A_1  - \frac{1}{2},2B_1  - \frac{1}{2}} \right)} \left( {\cos 2\vartheta } \right).
\ee
\normalsize
The functions Eq.{\,}\eqref{kgt-71} are the physically acceptable solutions of the Eq.{\,}\eqref{kg-4} (in the interval $0\leq \vartheta \leq \pi /2 $ ). At fixed $E$, the normalization and orthogonality of these Jacobi functions are specified by
{\fontsize{10pt}{9pt}\selectfont
\[
\int\limits_0^{\pi /2} {\Theta _k \left( \vartheta  \right)\Theta _{k'} \left( \vartheta  \right)\sin \vartheta d\vartheta }  = \delta _{k,k'},
\]
}
\normalsize
one obtains that
{\fontsize{10pt}{9pt}\selectfont
\be
\label{kgt-11}
N_k  = \sqrt {\frac{{2\left( {2k + 2A_1  + 2B_1} \right)\Gamma \left( {k + 2A_1  + 2B_1} \right)k!}}{{\Gamma \left( {k + 2A_1  + \frac{1}{2}} 
\right)
\Gamma \left( {k + 2B_1  + \frac{1}{2}} \right)}}} .
\ee
}
\normalsize
When $\nu=\nu '=0 \; (B_1=\frac{1}{2})$, the functions Eq.{\,}\eqref{kgt-71} coincide with the odd (anti-symmetrical) solutions Eq.{\,}\eqref{kgh-8}, i.e., $\Theta _k \left( \vartheta  \right)=\sqrt{2}\,{\rm H}_{2k+1} \left( \vartheta \right)$ on the hemisphere (up to phase) due the identity{\,}\cite{jafarov2024_qso_pdm}:
{\fontsize{10pt}{9pt}\selectfont
\[
P_{2k+1}^{\left( \tilde \alpha,\tilde \alpha \right)} \left( \cos \vartheta \right)=\frac{k!}{(2k+1)!}\cdot \frac{\Gamma (2k+\tilde \alpha +2)}{\Gamma (k+\tilde \alpha +1)} \cos \vartheta P_{k}^{\left( \tilde \alpha,\frac{1}{2}\right)} \left( \cos 2 \vartheta \right).
\]
}
\normalsize
One obtains the following value of the parameter $\Lambda=L'\left(L'+1\right)$ from the above expressions Eq.{\,}\eqref{kgt-6} and Eq.{\,}\eqref{kgt-7} of the parameter $\mu$:
{\fontsize{10pt}{9pt}\selectfont
\be
\label{kgt-8}
\Lambda \equiv \Lambda_k = 4\left( {k + A_1  + B_1 } \right)^2  - \frac{1}{4} - \gamma \left( {\beta ' + \nu '} \right) = L'_k \left( {L'_k  + 1} 
\right),
\ee
}
\normalsize
where the generalized orbital momentum $L'_k$  equals to
{\fontsize{10pt}{9pt}\selectfont
\be
\label{kgt-9}
L'_k  =\sqrt{\left( {2k + 2A_1  + 2B_1 } \right)^2 - \gamma \left( {\beta ' + \nu '} \right)}-\frac{1}{2}.
\ee
}
\normalsize
As one could expect, at $\beta=\beta'=0$ and $\nu=\nu'=0$ we have $L'_k  = 2k + \left| m \right| + 1$, which selects odd $l-|m|$, not necessarily odd $l$. Indeed, at $ \nu = \nu'= 0 \; (2B_1=1) $ the generalized orbital momentum Eq.{\,}\eqref{kgt-9} becomes equal to $L'_k  =\sqrt{\left( {2k + 1 + 2A_1 } \right)^2 - \gamma \beta '}-\frac{1}{2}$, i.e., only those values of quantum number Eq.{\,}\eqref{kgh-96} that correspond to odd states are realized.

\subsection{Zero-barrier limit and reflection parity}
\noindent
For the chosen hemisphere domain,
\begin{align}
\Lambda_k^{\mathrm{half}}&=(2k+a+b+1)^2-\tfrac14-\gamma(\beta'+\nu'),\\
\Lambda_j^{\mathrm{full}}\big|_{\nu=\nu'=0}&=(j+a+\tfrac12)^2-\tfrac14-\gamma\beta'.
\label{eq:domain_spectra_revision}
\end{align}
Taking $\nu,\nu'\to0$ while retaining the equatorial boundary gives $b\to1/2$ and $\Lambda_k^{\mathrm{half}}\to\Lambda_{2k+1}^{\mathrm{full}}$. This is the odd branch under equatorial reflection of the regular full-domain problem. A half-domain state itself has no reflection parity. In the disconnected two-sector system, both $(\Theta_L+\Theta_R)/\sqrt2$ and $(\Theta_L-\Theta_R)/\sqrt2$ are possible reflection eigenstates with identical eigenvalues. In the retained-wall zero-barrier limit the even combination generally has a derivative cusp and is not in the regular full-domain operator domain. For normalized functions the relation on $0<\vartheta<\pi/2$ is $\Theta_k^{\mathrm{half}}=\sqrt2\,H_{2k+1}$, up to phase.

When all angular couplings vanish, the hemisphere sequence is $l=2k+|m|+1$, whereas the regular full-domain sequence is $l=j+|m|$. The restriction is that $l-|m|$ is odd, not that $l$ itself must be odd. Eq.{\,}\eqref{eq:domain_spectra_revision} makes the two different boundary-domain limits explicit.

Substitution of Eq.{\,}\eqref{kgt-8} into Eq.{\,}\eqref{e-sp} yields the following expression for the energy spectrum:
{\fontsize{9pt}{9pt}\selectfont
\be
\label{e-spf}
\begin{split}
E^2  - M^2 c^4  &= \hbar ^2 c^2 \gamma D\left( {1 - \sigma _0 } \right)^2 \\
&\quad - \hbar ^2 c^2 \alpha ^2 \left[ {\frac{{p\sigma _0 \left( {1 - \sigma _0 } \right) - L'_k \left(L'_k+1 \right) -n^2 - \left( {2n + 1} \right)B}}{n + B}} \right]^2 .
\end{split}
\ee
}
\normalsize
Thus, the total wavefunctions for the considered quantum system with the double ring-shaped hyperbolic potential Eq.{\,}\eqref{new-drsho} are
{\fontsize{9.1pt}{9pt}\selectfont
\be
\label{wf-t}
\begin{split}
&\psi \left( {\vec r} \right) = C_n N_k \Phi_m(\varphi)\frac{1}{r}e^{ - 2\alpha r\sqrt {c_0 } } \left( {1 - e^{ - 2\alpha r} } \right)^{\frac{1}{2} + \sqrt {c_0  + c_2  - c_1 } }P_n^{\left( {2\sqrt {c_0 } ,2\sqrt {c_0  + c_2  - c_1 } } \right)} \\
&\times \left( {1 - 2e^{ - 2\alpha r} } \right) \left( {\sin \vartheta } \right)^{2A_1  - \frac{1}{2}} \left( {\cos \vartheta } \right)^{2B_1 } P_k^{\left( {2A_1  - \frac{1}{2},2B_1  - \frac{1}{2}} \right)} \left( {\cos 2\vartheta } \right).
\end{split}
\ee
}
\normalsize
\begin{figure}[tbp]
\centering
\includegraphics[width=\linewidth]{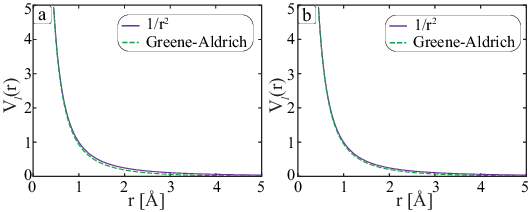}
\caption{\textbf{Exact and Greene--Aldrich centrifugal factors.} The functions $r^{-2}$ and $f_{\mathrm{GA}}(r)=\alpha^2/\sinh^2(\alpha r)$ are compared for \textbf{(a)} $\alpha=0.4588\,\text{\AA}^{-1}$ and \textbf{(b)} $\alpha=0.3848\,\text{\AA}^{-1}$, corresponding to the NaH and Na$_2$ radial parametrizations, respectively. Radial distances are expressed in \AA\ and both factors in $\text{\AA}^{-2}$. The approximation reproduces the leading $r^{-2}$ singularity as $r\to0$ but decays exponentially at
large $r$. This pointwise comparison is complemented by the state-resolved spectral tests in SM Fig.{\,}{S1} and
Fig.{\,}\ref{fig4}.}\label{fig2}
\end{figure}

\section{Non-relativistic limit $c \to \infty$}\label{sec3}
\noindent
In this section, we find the non-relativistic limits of the energy Eq.{\,}\eqref{e-spf} and radial wave function Eq.{\,}\eqref{wf-t} of the system under consideration and compare them with the corresponding results in Ref.{\,}\cite{dong2007}. In doing so, we note that the KG Eq.{\,}\eqref{kg-1} reduces to the Schr\"odinger equation for the full potential $V(r,\vartheta)$. First of all, one computes possible non-relativistic limits. These computations yield:
{\fontsize{10pt}{9pt}\selectfont
\be
\label{r-nr-lim}
\begin{split}
&\lim_{c \to \infty} \frac{\varepsilon^2}{4\alpha^2} = \frac{\mu E_N}{2\hbar^2\alpha^2}, \quad
\lim_{c \to \infty} \gamma = \gamma_N = \frac{2\mu}{\hbar^2}, \\
&\lim_{c \to \infty} p = p_N = 4\kappa, \quad
\lim_{c \to \infty} A = A_N, \\
&\lim_{c \to \infty} B = B_N = \delta + 1, \quad
\lim_{c \to \infty} (E - Mc^2) = E_N > 0.
\end{split}
\ee
}
\normalsize
Here, we used the following notations:
{\fontsize{9pt}{9pt}\selectfont
\begin{equation}
\begin{split}
\kappa &= \frac{\mu D}{2 \hbar^2 \alpha^2}, \quad
\\
A_N &= \sqrt{-\frac{\mu E_N}{2\hbar^2\alpha^2}+\kappa(1-\sigma_0)^2}, \quad
\delta = \tfrac12 \bigl[\sqrt{16\,\kappa\,\sigma_0^2 + (1 + 2L_{Nk})^2} - 1\bigr].
\end{split}
\end{equation}
}
\normalsize
where $L_{Nk}$ is obtained from $L'_k$ by replacing $\gamma$ with $\gamma_N$. Now it is easy to calculate the non-relativistic limit of energy equation Eq.{\,}\eqref{e-spf}, which yields:
{\fontsize{9.5pt}{9pt}\selectfont
\be
\label{en-spf}
\begin{split}
&E_N = D\left( {1 - \sigma _0 } \right)^2  \\
&- \frac{{ \hbar ^2 \alpha ^2 }}{2 \mu }\left[ {\frac{{4\kappa \sigma _0 \left( 1 - \sigma _0 \right) - L_{Nk} \left( L_{Nk}  + 1 \right) - \left( n + 1 \right)^2  - \left( 2n + 1 \right)\delta }}{{n + \delta  + 1}}} \right]^2 .
\end{split}
\ee
}
\normalsize

We drop the calculation of similar non-relativistic limit for the wavefunction Eq.{\,}\eqref{wf-t}, but, one needs to highlight that calculation of such a limit is not difficult and completely recovers known non-relativistic expression of the wavefunction.
\begin{figure*}[tbp]
\centering
\includegraphics[width=\linewidth]{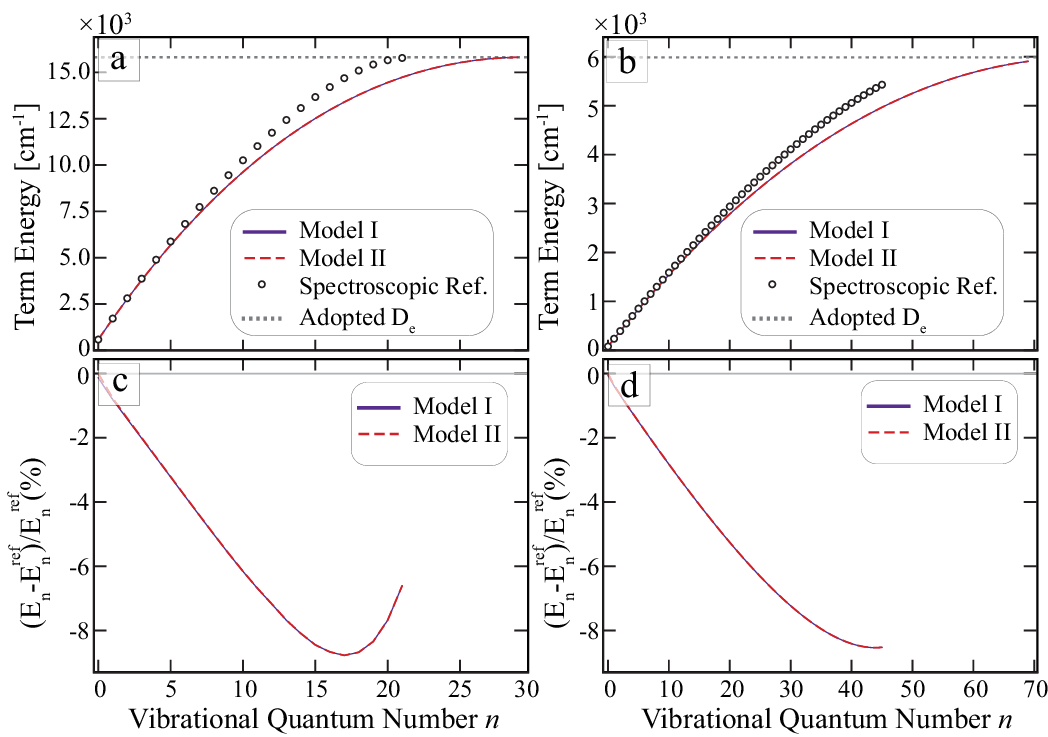}
\caption{\textbf{Vibrational term energies and residuals for NaH and Na$_2$.} Panels \textbf{(a,b)} compare Model I (solid purple curves), Model II (dashed red curves), and spectroscopic reference term values $E_n^{\mathrm{ref}}$ (open black circles) for NaH and Na$_2$, respectively. Term energies are measured from the radial potential minimum and displayed in $10^3\,\mathrm{cm}^{-1}$.
Dotted lines indicate the adopted dissociation energies, $D_e=15815\,\mathrm{cm}^{-1}$ for NaH and $5988\,\mathrm{cm}^{-1}$ for Na$_2$. Panels \textbf{(c,d)} show the percentage residuals $100\cdot(E_{n}-E_{n}^{\mathrm{ref}})/E_n^{\mathrm{ref}}$
calculated from Tables S1--S2. Model I uses $\Lambda=0$; Model II uses the fitted radial coefficients $\Lambda=0.201585$ and $0.491$ for NaH and Na$_2$, respectively. These fitted coefficients describe a radial sensitivity test and are not angular eigenvalues of the repulsive double-ring hemisphere problem. Model energies without reference values appear only in the upper panels.}\label{fig3}
\end{figure*}

\section{Results and Discussion}\label{sec4}
\noindent
The angular interaction can be rewritten as:
\begin{equation}
\frac{f(\vartheta)}{r^2}=\frac{\beta+\beta'}{\rho^2}+\frac{\nu+\nu'}{z^2}-\frac{\beta'+\nu'}{r^2},\qquad \rho=r\sin\vartheta,\quad z=r\cos\vartheta.
\label{eq:cylindrical_revision}
\end{equation}
Thus, the singular sets are the polar axis and the equatorial plane, as illustrated in Fig.{\,}\ref{fig1}. This is an idealized model of anisotropic confinement, without a microscopic derivation for an external environment. It is not an intrinsic interaction of an isolated, freely rotating diatomic molecule. Its four coefficients enter through three independent angular combinations. The molecular comparison below therefore tests the central radial potential with all angular couplings zero and the regular full angular domain.
\subsection{Molecular parametrization and provenance}
\noindent
The molecular illustration uses the non-relativistic spectrum in Eq.{\,}\eqref{en-spf}. With all angular couplings zero, the regular full-domain $l=0$ branch gives $\Lambda=0$; it must not be obtained by retaining a Dirichlet equatorial wall. Equal scalar--vector KG coupling is a solvable theoretical setting, not a fitted microscopic relativistic interaction for NaH or Na$_2$.

For an unambiguous Lambert-$W$ mapping{\,}\cite{wang2012}, write the spectroscopic wavenumber as $\widetilde\nu_e$, the ordinary frequency as $c\widetilde\nu_e$, and $D_e$ as an energy. Define
\begin{align}
q&=2\pi c\widetilde\nu_e\sqrt{\frac{\mu}{2D_e}},\\
\alpha&=\frac{qr_e+W_0(-qr_e e^{-qr_e})}{2r_e},\qquad qr_e>1.
\label{eq:alpha_Lambert}
\end{align}
Here, $W_0$ denotes the principal real branch of the Lambert $W$ function. For $qr_e>1$, this branch gives $\alpha>0$, whereas the alternative real branch $W_{-1}$ yields the zero-$\alpha$ root of the rearranged equation. The remaining parameters are $\sigma_0=\tanh(\alpha r_e)$ and $D=D_e/(1-\sigma_0)^2$.

The values used in the comparison are $D_e=15815\ \mathrm{cm}^{-1}$, $r_e=1.8870\ \text{\AA}$, $\widetilde\nu_e=1171.759\ \mathrm{cm}^{-1}$, and printed $\alpha=0.4588\ \text{\AA}^{-1}$ for NaH; and $D_e=5988\ \mathrm{cm}^{-1}$, $r_e=3.0785\ \text{\AA}$, $\widetilde\nu_e=159.109\ \mathrm{cm}^{-1}$, and printed $\alpha=0.3848\ \text{\AA}^{-1}$ for Na$_2$. Here and in the tables, energies in $\mathrm{cm}^{-1}$ denote energy divided by $hc$. Huang \textit{et al.}{\,}\cite{huang2010} report $D_e=15815\pm5\ \mathrm{cm}^{-1}$ for NaH; their Table III supplies the spectroscopically derived rotationless term values used here. The adopted Na$_2$ well depth is the historical value of Kusch and Hessel{\,}\cite{kusch1978}. The compiled constants do not define a new spectroscopic determination. The original table values are preserved in the SM; the independent approximation benchmark explicitly fixes its own rounded-input convention.

\subsection{Central radial comparison and its limitations}
\noindent
Fig.{\,}\ref{fig3} redraws the tabulated term values. Model I has $\Lambda=0$. Model II adds a fitted positive $K\Lambda/r^2$ contribution, evaluated with the GA factor, to align the ground level; it is retained only as a phenomenological sensitivity test. For a freely rotating molecule, the physical centrifugal coefficient is $l(l+1)$, not a continuously fitted number. Moreover, with nonnegative angular couplings and the retained equatorial Dirichlet boundary, the lowest angular separation eigenvalue obeys $\Lambda\geq2$. The values $0.201585$ and $0.491$ therefore cannot be eigenvalues of the stated repulsive double-ring hemisphere problem. Their ground-level agreement is fitted and is not independent evidence for angular confinement.

Using the tabulated Model I values, the relative deviations at $n=0,1,2,3$ are approximately $-0.13\%,-0.81\%,-1.41\%,-2.01\%$ for NaH and $-0.06\%,-0.39\%,-0.67\%,-0.96\%$ for Na$_2$. Thus the low-level accuracy must be stated quantitatively rather than extended without qualification to all intermediate levels. The maximum absolute percentage deviations within the supplied reference sets are approximately $8.77\%$ (NaH, $n=17$) and $8.53\%$ (Na$_2$, $n=44$). These tabulated term energies are measured from the potential minimum. At a fixed adopted $D_e$, their underestimation corresponds to overestimation of the remaining binding energy.

For Model I there is no centrifugal approximation at all, so the discrepancies arise from the radial model and its parametrization/reference-data choices. For Model II the independently bounded GA error is far smaller than the molecular residuals. Consequently these residuals must not be attributed mainly to the GA approximation. Matching the depth, minimum and curvature does not determine the global well shape or dissociation tail.

\begin{figure*}[tbp]
\centering
\includegraphics[width=\linewidth]{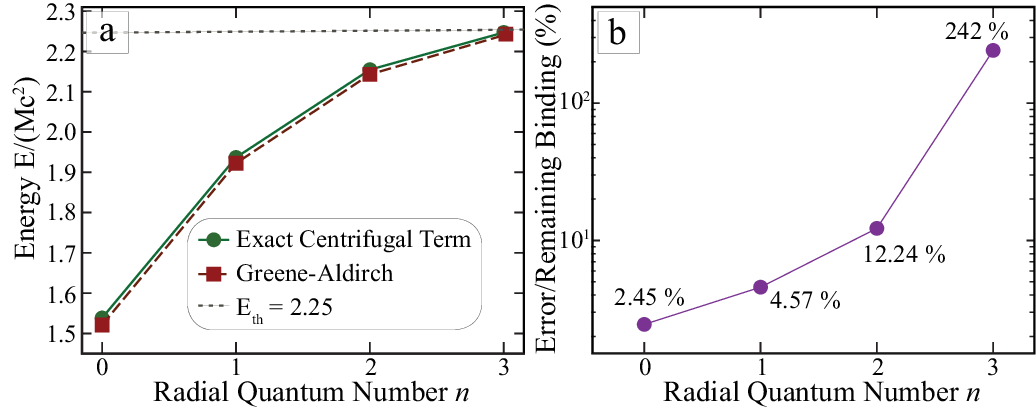}
\caption{\textbf{Relativistic bound-state benchmark for the centrifugal approximation.} \textbf{(a)} Bound-state energies obtained numerically with the unapproximated centrifugal term (green circles, solid line) and from the Greene--Aldrich quantization condition (red squares, dashed line). In units $\hbar=c=M=1$, the parameters are $D=5$, $\sigma_0=0.5$, $\alpha=0.2$, $\beta=0.05$, $\nu=0.1$, $\beta'=\nu'=0$, and $m=k=0$. The energy-dependent quantities $\gamma(E)$ and $\Lambda(E)$ are evaluated at each trial energy. The dotted line marks the continuum threshold $E_{\mathrm{th}}=2.25$. \textbf{(b)} Percentage error relative to the remaining binding energy, $100\cdot(E_n^{\mathrm{ex}}-E_n^{\mathrm{GA}})/(E_{\mathrm{th}}-E_n^{\mathrm{ex}})$, displayed on a logarithmic scale. Here, $E_n^{\mathrm{ex}}$ denotes the converged numerical reference energy obtained with the unapproximated centrifugal term.}\label{fig4}
\end{figure*}

The model predicts 30 NaH and 79 Na$_2$ bound vibrational levels in the stated central parametrization. The NaH result is a substantive limitation: Huang \textit{et al.}{\,}\cite{huang2010} infer $n=21$ as the highest bound vibrational level, corresponding to 22 levels, from their spectra and near-dissociation extrapolation. The additional model levels cannot simply be described as not yet observed. The Na$_2$ reference term values included in Table{\,}S2 extend to $n=45$, where $E_{45}^{\mathrm{ref}}=5428.611\,\mathrm{cm}^{-1}$. This level lies $559.389\,\mathrm{cm}^{-1}$ below the adopted dissociation threshold,
$D_e=5988\,\mathrm{cm}^{-1}$. The endpoint of the reference dataset therefore does not identify the highest physical bound vibrational level. The molecular comparisons assess the radial model over the available reference levels, with Na$_2$ percentage
residuals reaching approximately $8.53\%$. They do not provide an empirical test of the singular angular interaction.

\subsection{State-resolved validation of the radial approximation}\label{sec:validation}
\noindent
Write the Greene--Aldrich (GA) factor as $f_{\rm GA}(r)=\alpha^2/\sinh^2(\alpha r)$. The original point-wise comparison is retained in Fig.{\,}\ref{fig2}. Its fractional defect is
\begin{equation}
\eta(r)=1-\left[\frac{\alpha r}{\sinh(\alpha r)}\right]^2.
\label{eq:eta_revision}
\end{equation}
Localization near $r_e$ alone does not ensure $\eta\ll1$: the printed parameters give $\alpha r_e=0.8658$ for NaH and $1.1846$ for Na$_2$, with $\eta(r_e)=21.64\%$ and $36.09\%$, respectively. Nevertheless, the absolute spectral error may be small when the centrifugal energy itself is small.

For the non-relativistic problem define $K=\hbar^2/(2\mu)$ in consistent energy--length units and
\begin{equation}
H_{\rm ex}=-K\partial_r^2+V(r)+\frac{K\Lambda}{r^2},\quad
H_{\rm GA}=-K\partial_r^2+V(r)+K\Lambda f_{\rm GA}(r).
\end{equation}
Both operators use the same radial domain and parameters. We compare the $n^{th}$ bound levels using
\begin{equation}
\Delta_n=E_n^{\rm ex}-E_n^{\rm GA},\qquad
R_n=\frac{\Delta_n}{D_e-E_n^{\rm ex}},\qquad
\epsilon_n=\frac{\langle r^{-2}-f_{\rm GA}\rangle_n}{\langle r^{-2}\rangle_n}.
\label{eq:metrics_revision}
\end{equation}
The expectation in $\epsilon_n$ uses the normalized GA radial eigenfunction. $R_n$ is defined only for an exact bound state. For $\Lambda\geq0$, a bounded potential difference and the min--max principle give
\begin{equation}
0\leq\Delta_n\leq\frac{K\Lambda\alpha^2}{3},
\label{eq:bound_revision}
\end{equation}
for corresponding discrete levels. The proof and numerical convergence data are supplied in the SM. This nonrelativistic bound is not applied to the energy-dependent KG pencil.

A second-order finite-difference discretization and Richardson extrapolation were used, with analytic GA eigenvalues as an independent check. The same-grid exact-minus-GA difference reduces common discretization error. The molecular benchmark fixes the printed $\alpha,r_e,D_e$ values and uses $\sigma_0=\tanh(\alpha r_e)$, $D=D_e/(1-\sigma_0)^2$, and the isotope masses listed in the SM. It is a separately specified numerical benchmark and does not claim to reconstruct the unreported full-precision inputs behind the original molecular tables.

SM Fig.{\,}S1 presents the state-resolved nonrelativistic errors. For the diagnostic fitted values $\Lambda=0.201585$ (NaH) and $0.491$ (Na$_2$), $\Delta_0$ is approximately $0.21292$ and $0.02736\ \mathrm{cm}^{-1}$. The corresponding bounds are $0.24696$ and $0.03554\ \mathrm{cm}^{-1}$. At the highest bound model levels, $n=29$ and $78$, the absolute shifts decrease to approximately $0.03396$ and $0.00370\ \mathrm{cm}^{-1}$, but $100R_n$ rises to about $1.70\%$ and $1.06\%$. For $\Lambda=0$, the two operators coincide exactly. Increasing $\Lambda$ to 20 gives a stronger test: the NaH GA model predicts 30 bound states, whereas the exact radial operator has 29. The extra finite-box eigenvalue is a continuum state and is excluded from bound-state error ratios.

Fig.{\,}\ref{fig4} gives a direct relativistic benchmark, which solves the unsquared radial spectral condition while recalculating $\gamma(E)$ and $\Lambda(E)$ at every trial energy. In units $\hbar=c=M=1$, choose $D=5$, $\sigma_0=0.5$, $\alpha=0.2$, $\beta=0.05$, $\nu=0.1$, $\beta'=\nu'=0$, $m=k=0$. The equatorial barrier is retained. The threshold is $E_{\rm th}=2.25$. For $n=0,1,2,3$, the exact energies are approximately $1.538349,1.936758,2.154662,2.247130$, versus GA values $1.520939,1.922456,2.142992,2.240187$. Thus the absolute error decreases, but the relative remaining-binding errors are approximately $2.45\%,4.57\%,12.24\%,242\%$. The last value means that the GA prediction substantially overestimates the small remaining binding; it does not mean a $242\%$ error in the total energy. These are numerical model benchmarks, not molecular relativistic predictions.

\section{Conclusions}\label{sec5}
\noindent
In this work, we have obtained exact analytical solutions of the angular equation and approximate analytical solutions of the radial KG equation for a non-central Schi{\"o}berg-type hyperbolic radial potential supplemented by double ring-shaped angular terms, assuming equal scalar and vector couplings. The radial equation was treated with the GA approximation, and both radial and angular wavefunctions were expressed in terms of Jacobi polynomials. The angular separation eigenvalue leads to an implicit quantization condition for the relativistic bound-state energies. For $\nu+\nu'>0$, the angular interaction is singular at $\vartheta=\pi/2$. Under the principal boundary condition adopted there, the full angular interval separates into two dynamically disconnected sectors. In the limit-circle regime this separation requires the stated boundary-domain choice and is not shared by every self-adjoint extension. Reflection symmetry maps one sector to the other, so they possess identical separation spectra. The second sector therefore supplies mirror-related partner states rather than a distinct set of energy eigenvalues.

The limit in which the equatorial coupling approaches zero while the half-domain boundary condition is retained selects the odd-reflection spectral branch of the regular full-domain problem. The two-sector system itself supports both reflection combinations; a single hemisphere has no intrinsic parity. This limiting procedure differs from solving the zero-barrier problem directly on the full interval, where both even and odd branches are present. This distinction clarifies the restricted recovery of the central-potential spectrum in the sector formulation. The non-relativistic limit was also obtained. For NaH and Na$_2$, the calculated central-radial vibrational term energies agree reasonably well with spectroscopic reference values for the lowest states, whereas deviations increase for highly excited levels. The present model therefore provides an analytical benchmark for singular angular confinement and a limited near-equilibrium molecular illustration. Neither a new general solution method nor a microscopic double-ring interaction for free diatomic molecules is claimed.

Direct radial calculations separate the error of the centrifugal replacement from the error of the molecular potential model. The fitted-$\Lambda$ ground-level shifts are only about $0.213$ and $0.0274\ \mathrm{cm}^{-1}$ for NaH and Na$_2$, whereas the model's spectroscopic residuals can reach approximately $8.8\%$. For $\Lambda=0$ the centrifugal approximation error vanishes identically. In the independent dimensionless KG example, the relative error in the remaining binding grows from approximately $2.45\%$ to $242\%$ over four bound states, despite decreasing absolute energy errors. Near-threshold levels therefore require the exact centrifugal term and state-specific convergence checks. Quantitative molecular use also requires a suitable radial potential and verified input constants; the NaH bound-state-count mismatch rules out a validated dissociation spectrum for the present parametrization.

\section*{Data Availability Statement}
\noindent
The comparison tables, independently generated radial benchmarks, convergence data are provided with the Supplementary Material. Spectroscopic reference values are attributed to the cited literature. The numerical benchmarks state their parameter and mass conventions explicitly.

\section*{Author Contribution Statement}
\noindent
S.M.N. and V.H.B. contributed to the conceptualization, methodology, and analytical development of the model and solutions. S.M.N. and V.H.B. performed the formal analysis and interpreted the results. V.H.B. prepared the figures and carried out the numerical computations. S.M.N. and V.H.B. wrote the manuscript and revised it critically. All authors read and approved the final manuscript.

\section*{Conflict of Interest}
\noindent
The authors declare no competing interests. All research has been carried out within an appropriate ethical framework.

\bibliography{biblo}

@book{davydov1965,
  author    = {A. S. Davydov},
  title     = {Quantum Mechanics},
  publisher = {Pergamon Press},
  address   = {Oxford / New York},
  year      = {1965},
  note      = {First edition},
  url       = {https://archive.org/details/davydov-quantum-mechanics},
}

@book{landau1979,
  author    = {L. D. Landau and E. M. Lifshitz},
  title     = {Quantum Mechanics: Non-Relativistic Theory},
  edition   = {3},
  publisher = {Pergamon Press},
  address   = {Oxford / London},
  year      = {1979},
  url       = {https://www.elsevier.com/books/quantum-mechanics/landau/978-0-08-020940-1},
}

@book{flugge1994,
  author    = {Siegfried Fl\"ugge},
  title     = {Practical Quantum Mechanics, Vol. 1},
  publisher = {Springer},
  address   = {Berlin / Heidelberg},
  year      = {1994},
  doi       = {10.1007/978-3-642-61995-3},
  url       = {https://link.springer.com/book/10.1007/978-3-642-61995-3},
}

@book{greiner2000,
  author    = {Walter Greiner},
  title     = {Relativistic Quantum Mechanics},
  edition   = {3},
  publisher = {Springer},
  address   = {Berlin / Heidelberg},
  year      = {2000},
  doi       = {10.1007/978-3-662-04275-5},
  url       = {https://link.springer.com/book/10.1007/978-3-662-04275-5},
}

@article{nagiyev2019,
  author    = {Nagiyev, Sh. M. and Ahmadov, A. I.},
  title     = {Exact solution of the relativistic finite-difference equation for the {C}oulomb plus a ring-shaped-like potential},
  journal   = {Int. J. Mod. Phys. A},
  year      = {2019},
  volume    = {34},
  pages     = {1950089},
  doi       = {10.1142/S0217751X19500891},
  url       = {https://doi.org/10.1142/S0217751X19500891}
}

@article{nagiyev2024,
  author    = {A. I. Ahmadov and Sh. M. Nagiyev and A. N. Ikot and V. A. Tarverdiyeva},
  title     = {Analytical solutions for the {K}lein–{G}ordon equation with combined exponential type and ring-shaped potentials},
  journal   = {Sci. Rep.},
  year      = {2024},
  volume    = {14},
  pages     = {5527},
  doi       = {10.1038/s41598-024-53650-8},
  url       = {https://www.nature.com/articles/s41598-024-53650-8}
}

@article{quesne1988,
  author    = {Christiane Quesne},
  title     = {A new ring-shaped potential and its dynamical invariance algebra},
  journal   = {J. Phys. A: Math. Theor.},
  year      = {1988},
  month     = {jul},
  publisher = {},
  volume    = {21},
  number    = {14},
  pages     = {3093--3103},
  doi       = {10.1088/0305-4470/21/14/010},
  url       = {https://doi.org/10.1088/0305-4470/21/14/010}
}

@article{berkdemir2006,
  author    = {Cüneyt Berkdemir and Ayşe Berkdemir and Jiaguang Han},
  title     = {Bound state solutions of the {S}chr\"odinger equation for modified {K}ratzer’s molecular potential},
  journal   = {Chem. Phys. Lett.},
  year      = {2006},
  volume    = {417},
  pages     = {326--329},
  doi       = {10.1016/j.cplett.2005.10.039},
  url       = {https://doi.org/10.1016/j.cplett.2005.10.039}
}

@article{fu2008,
  author    = {Cheng, Yan-Fu and Dai, Tong-Qing},
  title     = {Exact solutions of the {S}chr\"odinger equation for a new ring-shaped nonharmonic oscillator potential},
  journal   = {Int. J. Mod. Phys. A},
  year      = {2008},
  volume    = {23},
  pages     = {1919--1927},
  doi       = {10.1142/S0217751X08039621},
  url       = {https://doi.org/10.1142/S0217751X08039621}
}

@article{chen2005,
  author    = {C. Y. Chen and S. H. Dong},
  title     = {Exactly complete solutions of the Coulomb potential plus a new ring-shaped potential},
  journal   = {Phys. Lett. A},
  year      = {2005},
  volume    = {335},
  pages     = {374--382},
  doi       = {10.1016/j.physleta.2004.12.062},
  url       = {https://www.sciencedirect.com/science/article/abs/pii/S0375960104010518}
}

@article{mbadjoun2019,
author = {Mbadjoun, B. Tchana and Ema’a, J. M. Ema’a and Yomi, Jean and Abiama, P. Ele and Ben-Bolie, G. H. and Ateba, P. Owono},
title = {Factorization method for exact solution of the non-central modified {K}illingbeck potential plus a ring-shaped-like potential},
journal = {Mod. Phys. Lett. A},
volume = {34},
number = {10},
pages = {1950072},
year = {2019},
doi = {10.1142/S021773231950072X},
URL = {https://doi.org/10.1142/S021773231950072X}
}

@article{berkdemir2009,
  author    = {C. J. Berkdemir},
  title     = {A novel angle-dependent potential and its exact solution},
  journal   = {J. Math. Chem.},
  year      = {2009},
  volume    = {46},
  pages     = {139--154},
  doi       = {10.1007/s10910-008-9447-7},
  url       = {https://doi.org/10.1007/s10910-008-9447-7}
}

@article{ahmadov2013,
  author    = {H. I. Ahmadov and C. Aydin and N. Sh. Huseynova and O. Uzun},
  title     = {Analytical solutions of the {S}chr\"odinger equation with the {M}anning–{R}osen potential plus a ring-shaped-like potential},
  journal   = {Int. J. Mod. Phys. E},
  year      = {2013},
  volume    = {22},
  pages     = {1350072},
  doi       = {10.1142/S0218301313500729},
  url       = {https://www.worldscientific.com/doi/10.1142/S0218301313500729}
}

@article{zhang2010,
  author    = {M. C. Zhang and B. An and H. F. Guo-Qing},
  title     = {Exact solutions of a new {C}oulomb ring-shaped potential},
  journal   = {J. Math. Chem.},
  year      = {2010},
  volume    = {48},
  pages     = {876--882},
  doi       = {10.1007/s10910-010-9715-1},
  url       = {https://link.springer.com/article/10.1007/s10910-010-9715-1}
}

@article{gonul2000,
  author    = {B. G\"on\"ul and I. Zorba},
  title     = {Supersymmetric solutions of non-central potentials},
  journal   = {Phys. Lett. A},
  volume    = {269},
  number    = {2},
  pages     = {83--88},
  year      = {2000},
  issn      = {0375-9601},
  doi       = {https://doi.org/10.1016/S0375-9601(00)00252-8},
  url       = {https://www.sciencedirect.com/science/article/pii/S0375960100002528}  
}

@article{dutra2006,
  author    = {A. de Souza Dutra and G. Chen},
  title     = {On some classes of exactly-solvable {K}lein–{G}ordon equations},
  journal   = {Phys. Lett. A},
  year      = {2006},
  volume    = {349},
  pages     = {297--301},
  doi       = {10.1016/j.physleta.2005.09.056},
  url       = {https://www.sciencedirect.com/science/article/abs/pii/S0375960105008293}  
}

@article{rojas2005,
  title     = {Scattering of a {K}lein-{G}ordon particle by a {W}oods-{S}axon potential},
  author    = {Rojas, Clara and Villalba, V\'{\i}ctor M.},
  journal   = {Phys. Rev. A},
  volume    = {71},
  issue     = {5},
  pages     = {052101},
  numpages  = {4},
  year      = {2005},
  month     = {May},
  publisher = {American Physical Society},
  doi       = {10.1103/PhysRevA.71.052101},
  url       = {https://link.aps.org/doi/10.1103/PhysRevA.71.052101}, 
}

@article{adame1989,
  author    = {F. Dom{\'\i}nguez-Adame},
  title     = {Bound states of the {K}lein--{G}ordon equation with vector and scalar {H}ulth{\'e}n-type potentials},
  journal   = {Phys. Lett. A}, 
  volume    = {136},
  number    = {4},
  pages     = {175--177},
  year      = {1989},
  issn      = {0375-9601},
  doi       = {10.1016/0375-9601(89)90555-0},
  url       = {https://www.sciencedirect.com/science/article/pii/0375960189905550}
}

@article{zhang2005,
  author    = {Zhang Xue-Ao and Chen Ke and Duan Zheng-Lu},
  title     = {Bound states of {K}lein–{G}ordon equation and Dirac equation for ring-shaped non-spherical oscillator scalar and vector potentials},
  journal   = {Chinese Phys.},
  year      = {2005},
  month     = {jan},
  volume    = {14},
  number    = {1},
  pages     = {42--44},
  doi       = {10.1088/1009-1963/14/1/009},
  url       = {https://doi.org/10.1088/1009-1963/14/1/009}
}

@article{lu2005,
  author    = {Lu Fa-Lin and Chen Chang-Yuan and Sun Dong-Sheng},
  title     = {Bound states of {K}lein–{G}ordon equation for double ring-shaped oscillator scalar and vector potentials},
  journal   = {Chinese Phys.},
  year      = {2005},
  volume    = {14},
  number    = {3},
  pages     = {463--467},
  doi       = {10.1088/1009-1963/14/3/005},
  url       = {https://doi.org/10.1088/1009-1963/14/3/005}
}

@article{dong2004,
  author = {Shi-Hai Dong and Guo-Hua Sun and M. Lozada-Cassou},
  title = {An algebraic approach to the ring-shaped non-spherical oscillator},
  journal   = {Phys. Lett. A},
  volume = {328},
  number = {4},
  pages = {299-305},
  year = {2004},
  issn = {0375-9601},
  doi = {https://doi.org/10.1016/j.physleta.2004.06.037},
  url = {https://www.sciencedirect.com/science/article/pii/S0375960104008357},
}

@article{villalba2006,
  author    = {Villalba, V\'{i}ctor M. and Rojas, Clara},
  title     = {Bound states of the {K}lein–{G}ordon equation in the presence of short range potentials},
  journal   = {Int. J. Mod. Phys. A},
  volume    = {21},
  number    = {02},
  pages     = {313--325},
  year      = {2006},  
  doi       = {10.1142/S0217751X06025158},
  url       = {https://doi.org/10.1142/S0217751X06025158}  
}

@article{badalov2023,
  author    = {Badalov, V H and Badalov, S V},
  title     = {Generalised tanh-shaped hyperbolic potential: {K}lein–{G}ordon equation's bound state solution},
  journal   = {Commun. Theor. Phys.},
  year      = {2023},
  volume    = {75},
  number    = {7},
  pages     = {75003},
  doi       = {10.1088/1572-9494/acd441},
  url       = {https://doi.org/10.1088/1572-9494/acd441}
}

@article{yasuk2006,
  author = {Yasuk, F. and Durmus, A. and Boztosun, I.},
  title = {Exact analytical solution to the relativistic {K}lein-{G}ordon equation with noncentral equal scalar and vector potentials},
  journal   = {J. Math. Phys.},
  volume = {47},
  number = {8},
  pages = {082302},
  year = {2006},
  month = {08},
  doi = {10.1063/1.2227258},
  url = {https://doi.org/10.1063/1.2227258}  
}

@article{yi2004,
  author    = {L. Z. Yi and Y. F. Diao and J. Y. Liu and C. S. Jia},
  title     = {Bound states of the {K}lein–{G}ordon equation with vector and scalar {R}osen–{M}orse-type potentials},
  journal   = {Phys. Lett. A},
  year      = {2004},
  volume    = {333},
  pages     = {212},
  doi       = {10.1016/j.physleta.2004.10.054},
  url       = {https://www.sciencedirect.com/science/article/abs/pii/S0375960104003702}
}

@article{alhaidari2006,
  author    = {A. D. Alhaidari and H. Bahlouli and A. Al-Hasan},
  title     = {Dirac and {K}lein–{G}ordon equations with equal scalar and vector potentials},
  journal   = {Phys. Lett. A},
  year      = {2006},
  volume    = {349},
  pages     = {87},
  doi       = {10.1016/j.physleta.2005.09.008},
  url       = {https://www.sciencedirect.com/science/article/abs/pii/S0375960105014726}
}

@article{schioberg1986,
  author    = {D. Schi\"oberg},
  title     = {The energy eigenvalues of hyperbolical potential functions},
  journal   = {Mol. Phys.},
  year      = {1986},
  volume    = {59},
  pages     = {1123--1137},
  doi       = {10.1080/00268978600102631},
  url       = {https://www.tandfonline.com/doi/abs/10.1080/00268978600102631}
}

@article{jun2005,
  author    = {L. Jun and Q. Hui-Xian and L. Liang-Mei and L. Feng-Ling},
  title     = {Rotation and vibration of diatomic molecule oscillator with hyperbolic potential function},
  journal   = {Chinese Phys.},
  year      = {2005},
  volume    = {14},
  pages     = {2402--2406},
  doi       = {10.1088/1009-1963/14/12/005},
  url       = {https://cpb.iphy.ac.cn/EN/10.1088/1009-1963/14/12/005}
}

@article{dong2007,
  author    = {S. Dong and J. Garcia-Ravelo and S. H. Dong},
  title     = {Analytical approximations to the $l$-wave solutions of the {S}chrödinger equation with an exponential-type potential},
  journal   = {Phys. Scr.},
  year      = {2007},
  volume    = {76},
  pages     = {393--396},
  doi       = {10.1088/0031-8949/76/4/019},
  url       = {https://doi.org/10.1088/0031-8949/76/4/019}
}

@article{ikhdair2009,
  author    = {S. M. Ikhdair},
  title     = {Rotational and vibrational diatomic molecule in the {K}lein–{G}ordon equation with hyperbolic scalar and vector potential},
  journal   = {Int. J. Mod. Phys. C},
  year      = {2009},
  volume    = {20},
  pages     = {1563--1582},
  doi       = {10.1142/S0129183109014606},
  url       = {https://ideas.repec.org/a/wsi/ijmpcx/v20y2009i10ns0129183109014606.html}
}

@article{caprio1989,
  author    = {M. V. Carpio-Bernido and C. C. Bernido},
  title     = {An exact solution of a ring-shaped oscillator plus a $\csc^2\vartheta\, r^2$ potential},
  journal   = {Phys. Lett. A},
  year      = {1989},
  volume    = {134},
  pages     = {395--399},
  doi       = {10.1016/0375-9601(89)90357-5},
  url       = {https://www.sciencedirect.com/science/article/abs/pii/0375960189903575}
}

@article{aktas2009,
  author    = {M. Akta\c{s}},
  title     = {Exact bound state solutions of the {S}chr\"odinger equation for noncentral potential via the {N}ikiforov–{U}varov method},
  journal   = {Int. J. Theor. Phys.},
  year      = {2009},
  volume    = {48},
  pages     = {2154--2163},
  doi       = {10.1007/s10773-009-9993-1},
  url       = {https://link.springer.com/article/10.1007/s10773-009-9993-1}
}

@article{chen2013,
  author    = {Chang-Yuan Chen and Yuan You and Xiao-Hua Wang and Shi-Hai Dong},
  title     = {Exact solutions of the {S}chrödinger equation with double ring-shaped oscillator},
  journal   = {Phys. Lett. A},
  volume = {377},
  number = {23},
  pages = {1521--1525},
  year = {2013},
  issn = {0375-9601},
  doi       = {10.1016/j.physleta.2013.04.026},
  url       = {https://www.sciencedirect.com/science/article/pii/S0375960113003861}
}

@article{yasuk2008,
  author    = {Yasuk, F and Durmus, A},
  title     = {Relativistic solutions for double ring-shaped oscillator potential via asymptotic iteration method},
  journal   = {Phys. Scr.},
  year      = {2007},
  month     = {dec},
  publisher = {},
  volume    = {77},
  number    = {1},
  pages     = {015005},
  doi       = {10.1088/0031-8949/77/01/015005},
  url       = {https://doi.org/10.1088/0031-8949/77/01/015005}  
}

@article{maghsoodi2013,
  author    = {E. Maghsoodi and H. Hassanabadi and S. Zarrinkamar},
  title     = {Exact solutions of {D}irac equation with {P}\"oschl–{T}eller double-ring-shaped {C}oulomb potential via {N}ikiforov–{U}varov method},
  journal   = {Chinese Phys. B},
  year      = {2013},
  volume    = {22},
  number    = {3},
  pages     = {030302},
  doi       = {10.1088/1674-1056/22/3/030302},
  url       = {https://doi.org/10.1088/1674-1056/22/3/030302}  
}

@article{durmus2007,
  author    = {Durmus, Aysen and Yasuk, Fevziye},
  title     = {Relativistic and nonrelativistic solutions for diatomic molecules in the presence of double ring-shaped {K}ratzer potential},
  journal   = {J. Chem. Phys.},
  year      = {2007},
  volume    = {126},
  pages     = {074108},
  issn      = {0021-9606},
  doi       = {10.1063/1.2566432},
  url       = {https://doi.org/10.1063/1.2566432}  
}

@article{calogero1969,
  author    = {Calogero, F.},
  title     = {Solution of a three-body problem in one dimension},
  journal   = {J. Math. Phys.},
  year      = {1969},
  volume    = {10},
  pages     = {2191--2196},
  doi       = {10.1063/1.1664820},
  url       = {https://doi.org/10.1063/1.1664820}  
}

@article{pekeris1934,
  author    = {C. L. Pekeris},
  title     = {The {R}otation-{V}ibration {C}oupling in {D}iatomic {M}olecules},
  journal   = {Phys. Rev.},
  year      = {1934},
  volume    = {45},
  number    = {2},
  pages     = {98--103},
  doi       = {10.1103/PhysRev.45.98},
  url       = {https://link.aps.org/doi/10.1103/PhysRev.45.98}  
}

@book{koekoek2010,
  author    = {R. Koekoek and P. A. Lesky and R. F. Swarttouw},
  title     = {Hypergeometric {O}rthogonal {P}olynomials and {T}heir $q$-{A}nalogues},
  series    = {Springer Monographs in Mathematics},
  publisher = {Springer},
  address   = {Berlin / Heidelberg},
  year      = {2010},
  doi       = {10.1007/978-3-642-05014-5},
  url       = {https://link.springer.com/book/10.1007/978-3-642-05014-5}  
}

@article{wang2012,
  author    = {P.-Q. Wang and J.-Y. Liu and L.-H. Zhang and S.-Y. Cao and C.-S. Jia},
  title     = {Improved expressions for the {S}chi\"oberg potential energy models for diatomic molecules},
  journal   = {J. Mol. Spectrosc.},
  year      = {2012},
  volume    = {278},
  issn      = {0022-2852},
  pages     = {23--26},
  doi       = {10.1016/j.jms.2012.07.001},
  url       = {https://doi.org/10.1016/j.jms.2012.07.001}  
}

@article{huang2010,
  author    = {Huang, Hsien-Yu and Lu, Tsai-Lien and Whang, Thou-Jen and Chang, Yung-Yung and Tsai, Chin-Chun},
  title     = {Dissociation energy of the ground state of {N}a{H}},
  journal   = {J. Chem. Phys.},
  year      = {2010},
  volume    = {133},
  pages     = {044301},
  doi       = {10.1063/1.3458914},
  url       = {https://doi.org/10.1063/1.3458914}  
}

@article{kusch1978,
  author    = {P. Kusch and M. M. Hessel},
  title     = {An analysis of the $B\,^1\Pi_u$–$X\,^1\Sigma_g^+$ band system of {Na}$_2$},
  journal   = {J. Chem. Phys.},
  year      = {1978},
  volume    = {68},
  pages     = {2591--2606},
  doi       = {10.1063/1.436117},
  url       = {https://doi.org/10.1063/1.436117}  
}

@article{hartmann1972,
  author    = {H. Hartmann},
  title     = {Die {B}ewegung eines {K}{\"o}rpers in einem ringf{\"o}rmigen {P}otentialfeld},
  journal   = {Theor. Chim. Acta},
  year      = {1972},
  volume    = {24},
  pages     = {201--206},
  doi       = {10.1007/BF00641399},
  url       = {https://doi.org/10.1007/BF00641399},
  language  = {German}
}

@article{hautot1973,
  author    = {A. Hautot},
  title     = {Exact motion in noncentral electric fields},
  journal   = {J. Math. Phys.},
  year      = {1973},
  volume    = {14},
  number    = {10},
  pages     = {1320--1327},
  doi       = {10.1063/1.1666184},
  url       = {https://dx.doi.org/10.1063/1.1666184}
}

@article{jafarov2024_qso_pdm,
  author    = {E. I. Jafarov and S. M. Nagiyev},
  title     = {Quantum singular oscillator with potential controlled by position-dependent mass},
  journal   = {Turk. J. Phys.},
  year      = {2024},
  volume    = {48},
  number    = {6},
  pages     = {153--179},
  doi       = {10.55730/1300-0101.2768},
  url       = {https://doi.org/10.55730/1300-0101.2768}
}

@article{greene1976,
  author    = {R. L. Greene and C. Aldrich},
  title     = {Variational wave functions for a screened {C}oulomb potential},
  journal   = {Phys. Rev. A},
  year      = {1976},
  volume    = {14},
  number    = {6},
  pages     = {2363--2366},
  doi       = {10.1103/PhysRevA.14.2363},
  url       = {https://link.aps.org/doi/10.1103/PhysRevA.14.2363}
}

@book{abramowitz1964,
  editor    = {Milton Abramowitz and Irene A. Stegun},
  title     = {Handbook of {M}athematical {F}unctions with {F}ormulas, {G}raphs, and {M}athematical {T}ables},
  publisher = {Dover},
  address   = {New York},
  year      = {1964},
  edition   = {9$^{th}$ Dover printing, 10$^{th}$ GPO printing},
  url       = {https://personal.math.ubc.ca/~cbm/aands/},
}

@article{Sarathi_2025,
author = {Sarathi, Partha and Rawat, Bhaskar Singh},
title = {Exact solution of {S}chrödinger equation for the complex {M}orse potential to investigate physical systems with position-dependent complex mass},
year = {2025},
month = {aug},
publisher = {IOP Publishing},
volume = {100},
number = {8},
pages = {085259},
journal = {Phys. Scr.},
doi = {10.1088/1402-4896/adf94e},
url = {https://doi.org/10.1088/1402-4896/adf94e},
}

@article{Aid_2024,
author = {Aid, Salah Eddine and Boukabcha, Hocine and Bentridi, Salah Eddine},
title = {Path integral solution for a Dirac particle in a {G}eneralized {I}nverse {Q}uadratic {Y}ukawa potential},
year = {2024},
month = {aug},
publisher = {IOP Publishing},
volume = {99},
number = {9},
pages = {095408},
journal = {Phys. Scr.},
doi = {10.1088/1402-4896/ad6f54},
url = {https://doi.org/10.1088/1402-4896/ad6f54},
}

@article{Kadja_2024,
author = {Kadja, A},
title = {Complete solutions of the {D}irac equation with q-deformed hyperbolic {P}öschl-{T}eller potential plus a trigonometric {S}carf II potential},
year = {2024},
month = {jun},
publisher = {IOP Publishing},
volume = {99},
number = {7},
pages = {075405},
journal = {Phys. Scr.},
doi = {10.1088/1402-4896/ad5239},
url = {https://doi.org/10.1088/1402-4896/ad5239},
}

@article{de_Oliveira_2021,
author = {de Oliveira, M D and Schmidt, Alexandre G M},
title = {Quasi-exact solution of the {D}irac equation on curved space-time with {C}oulomb scalar and vector potentials and {M}ie-type tensor potential with pseudo-spin and spin symmetries},
year = {2021},
month = {feb},
publisher = {IOP Publishing},
volume = {96},
number = {5},
pages = {055301},
journal = {Phys. Scr.},
doi = {10.1088/1402-4896/abe495},
url = {https://doi.org/10.1088/1402-4896/abe495},
}

@article{Maghsoodi_2012,
author = {Maghsoodi, E and Hassanabadi, H and Aydoğdu, O},
title = {Dirac particles in the presence of the {Y}ukawa potential plus a tensor interaction in {SUSYQM} framework},
year = {2012},
month = {jul},
publisher = {IOP Publishing},
volume = {86},
number = {1},
pages = {015005},
journal = {Phys. Scr.},
doi = {10.1088/0031-8949/86/01/015005},
url = {https://doi.org/10.1088/0031-8949/86/01/015005},
}

@article{Boudjedaa_2024,
author = {Boudjedaa, Badredine and Ahmed, Faizuddin},
title = {Topological defects on solutions of the non-relativistic equation for extended double ring-shaped potential},
year = {2024},
month = {jul},
publisher = {IOP Publishing},
volume = {76},
number = {8},
pages = {085102},
journal = {Commun. Theor. Phys.},
doi = {10.1088/1572-9494/ad4c5e},
url = {https://doi.org/10.1088/1572-9494/ad4c5e},
}

@article{Ahmed_2023,
author = {Ahmed, Faizuddin},
title = {Klein–{G}ordon oscillator with magnetic and quantum flux fields in non-trivial topological space-time},
year = {2023},
month = {feb},
publisher = {IOP Publishing},
volume = {75},
number = {2},
pages = {025202},
journal = {Commun. Theor. Phys.},
doi = {10.1088/1572-9494/aca650},
url = {https://doi.org/10.1088/1572-9494/aca650},
}

@article{Oluwadare_2012,
author = {Oluwadare, O J and Oyewumi, K J and Akoshile, C O and Babalola, O A},
title = {Approximate analytical solutions of the relativistic equations with the Deng–Fan molecular potential including a {P}ekeris-type approximation to the (pseudo or) centrifugal term},
year = {2012},
month = {aug},
publisher = {IOP Publishing},
volume = {86},
number = {3},
pages = {035002},
author = {Oluwadare, O J and Oyewumi, K J and Akoshile, C O and Babalola, O A},
title = {Approximate analytical solutions of the relativistic equations with the Deng–Fan molecular potential including a Pekeris-type approximation to the (pseudo or) centrifugal term},
journal = {Phys. Scr.},
doi = {10.1088/0031-8949/86/03/035002},
url = {https://doi.org/10.1088/0031-8949/86/03/035002},
}

@article{gitman2010,
  author  = {Gitman, D. M. and Tyutin, I. V. and Voronov, B. L.},
  title   = {Self-adjoint extensions and spectral analysis in the {{C}alogero} problem},
  journal = {J. Phys. A: Math. Theor.},
  volume  = {43},
  number  = {14},
  pages   = {145205},
  year    = {2010},
  doi     = {10.1088/1751-8113/43/14/145205},
}

@article{gesztesy2024,
title = {On domain properties of {B}essel-type operators},
author = {Fritz Gesztesy and Michael M. H. Pang and Jonathan Stanfill},
journal = {Discrete Contin. Dyn. Syst. - S},
volume = {17},
number = {5&6},
pages = {1911--1946},
year = {2024},
issn = {1937-1632},
doi = {10.3934/dcdss.2022201},
url = {https://www.aimsciences.org/article/id/63968dad80e8bc15ce45ee4c}
}

@article{tsutsui2003,
title = {Connection conditions and the spectral family under singular potentials},
author = {Izumi Tsutsui and Tamás Fülöp and Taksu Cheon},
year = {2002},
month = {dec},
publisher = {},
volume = {36},
number = {1},
pages = {275},
journal = {J. Phys. A: Math. Gen.},
doi = {10.1088/0305-4470/36/1/319},
url = {https://doi.org/10.1088/0305-4470/36/1/319},
}

@article{feher2005,
  author  = {Feh{\'e}r, L. and Tsutsui, I. and F{\"u}l{\"o}p, T.},
title = {Inequivalent quantizations of the three-particle {{C}alogero} model constructed by separation of variables},
  journal = {Nucl. Phys. B},
  volume  = {715},
  number  = {3},
  pages   = {713--757},
  year    = {2005},
  doi     = {10.1016/j.nuclphysb.2005.02.006},
}

@article{sun2015parity,
  author  = {Sun, Dong-Sheng and Lu, Fa-Lin and You, Yuan and Chen, Chang-Yuan and Dong, Shi-Hai},
  title   = {Parity inversion property of the double ring-shaped oscillator in cylindrical coordinates},
  journal = {Mod. Phys. Lett. A},
  volume  = {30},
  number  = {39},
  pages   = {1550200},
  year    = {2015},
  doi     = {10.1142/S0217732315502004}
}

\end{document}